\documentclass[12pt]{article}
\usepackage[pass,paperwidth=8.5in,paperheight=11in,margin=1in]{geometry}
\usepackage{sectsty}
\usepackage{amssymb,amsmath}
\usepackage{ifxetex,ifluatex}
\usepackage{fixltx2e} 
\usepackage{graphicx,psfrag,epsf}
\usepackage{mathptmx}
\usepackage{enumerate}
\usepackage{algorithm}
\usepackage{algpseudocode}
\usepackage{enumitem}
\usepackage{bm}
\usepackage{natbib}
\usepackage{xcolor}
\usepackage{lmodern}
\usepackage{booktabs}
\usepackage{longtable}
\usepackage{url} 

\usepackage{titlesec}
\titlelabel{\thetitle.\quad}

\newcommand{\blind}{0}

\subsectionfont{\itshape}
\subsubsectionfont{\itshape}

\ifnum 0\ifxetex 1\fi\ifluatex 1\fi=0 
  \usepackage[T1]{fontenc}
  \usepackage[utf8]{inputenc}
  \usepackage{textcomp} 
\else 
  \usepackage{unicode-math}
  \defaultfontfeatures{Ligatures=TeX,Scale=MatchLowercase}
\fi
\IfFileExists{upquote.sty}{\usepackage{upquote}}{}
\IfFileExists{microtype.sty}{%
\usepackage[]{microtype}
\UseMicrotypeSet[protrusion]{basicmath} 
}{}
\IfFileExists{parskip.sty}{%
\usepackage{parskip}
}{
\setlength{\parindent}{0pt}
\setlength{\parskip}{6pt plus 2pt minus 1pt}
}
\usepackage{hyperref}
\hypersetup{
            pdftitle={Hamiltonian Monte Carlo in R},
            pdfauthor={Samuel Thomas and Wanzhu Tu},
            pdfborder={0 0 0},
            breaklinks=true}
\usepackage{color}
\usepackage{fancyvrb}

\DefineVerbatimEnvironment{Highlighting}{Verbatim}{commandchars=\\\{\}}
\usepackage{framed}
\definecolor{shadecolor}{RGB}{248,248,248}

\newcommand{\ControlFlowTok}[1]{\textcolor[rgb]{0.13,0.29,0.53}{\textbf{#1}}}
\newcommand{\DataTypeTok}[1]{\textcolor[rgb]{0.13,0.29,0.53}{#1}}
\newcommand{\DecValTok}[1]{\textcolor[rgb]{0.00,0.00,0.81}{#1}}

\newcommand{\FloatTok}[1]{\textcolor[rgb]{0.00,0.00,0.81}{#1}}

\newcommand{\KeywordTok}[1]{\textcolor[rgb]{0.13,0.29,0.53}{\textbf{#1}}}
\newcommand{\NormalTok}[1]{#1}
\newcommand{\OperatorTok}[1]{\textcolor[rgb]{0.81,0.36,0.00}{\textbf{#1}}}

\newcommand{\StringTok}[1]{\textcolor[rgb]{0.31,0.60,0.02}{#1}}

\newcommand{\pkg}[1]{{\normalfont\fontseries{b}\selectfont #1}}
\let\proglang=\textsf

\usepackage{graphicx,grffile}
\makeatletter
\def\maxwidth{\ifdim\Gin@nat@width>\linewidth\linewidth\else\Gin@nat@width\fi}
\def\maxheight{\ifdim\Gin@nat@height>\textheight\textheight\else\Gin@nat@height\fi}
\makeatother
\setkeys{Gin}{width=\maxwidth,height=\maxheight,keepaspectratio}
\ifx\paragraph\undefined\else
\let\oldparagraph\paragraph
\renewcommand{\paragraph}[1]{\oldparagraph{#1}\mbox{}}
\fi
\ifx\subparagraph\undefined\else
\let\oldsubparagraph\subparagraph
\renewcommand{\subparagraph}[1]{\oldsubparagraph{#1}\mbox{}}
\fi

\renewcommand\thesection{\arabic{section}}
\renewcommand\thesubsection{\thesection.\arabic{subsection}}

\begin{document}

\def\spacingset#1{\renewcommand{\baselinestretch}%
{#1}\small\normalsize} \spacingset{1}


\if0\blind
{
  \title{\bf Learning Hamiltonian Monte Carlo in R}
  \author{Samuel Thomas and Wanzhu Tu\thanks{
    Department of Biostatistics and Health Data Science, Indiana University School of Medicine, Email: \texttt{wtu1@iu.edu}}\hspace{.2cm}\\
    Indiana University School of Medicine}
  \maketitle
} \fi

\if1\blind
{
  \bigskip
  \bigskip
  \bigskip
  \begin{center}
    {\LARGE\bf Title}
\end{center}
  \medskip
} \fi

\bigskip
\begin{abstract}
Hamiltonian Monte Carlo (HMC) is a powerful tool for Bayesian computation. In comparison with the traditional Metropolis-Hastings algorithm, HMC offers greater computational efficiency, especially in higher dimensional or more complex modeling situations. To most statisticians, however, the idea of HMC comes from a less familiar origin, one that is based on the theory of classical mechanics. Its implementation, either through \proglang{Stan} or one of its derivative programs, can appear opaque to beginners. A lack of understanding of the inner working of HMC, in our opinion, has hindered its application to a broader range of statistical problems. In this article, we review the basic concepts of HMC in a language that is more familiar to statisticians, and we describe an HMC implementation in \proglang{R}, one of the most frequently used statistical software environments. We also present  \pkg{hmclearn}, an \proglang{R} package for learning HMC. This package contains a general-purpose HMC function for data analysis. We illustrate the use of this package in common statistical models. In doing so, we hope to promote this powerful computational tool for wider use. Example code for common statistical models is presented as supplementary material for online publication.
\end{abstract}

\noindent%
{\it Keywords:}  Bayesian computation, Hamiltonian Monte Carlo, MCMC, Stan
\vfill

\newpage
\spacingset{1.45} 
\section{Introduction}
\label{sec:intro}

Hamiltonian Monte Carlo (HMC) is one of the newer Markov Chain Monte Carlo (MCMC) methods for Bayesian computation. An essential advantage of HMC over the traditional MCMC methods, such as the Metropolis-Hastings algorithm, is its greatly improved computational efficiency, especially in higher-dimensional and more complex models. But despite the method's computational prowess and the existence of excellent introductions \citep{nealhandbook2011,betancourtconceptual2017}, practitioners still face daunting challenges in applying the method to their own applications. Difficulties mainly arise in three areas: (1) unfamiliarity with the theory behind the algorithm, (2) lack of understanding of how the existing software works, (3) inability to tune the HMC parameters. These difficulties have limited the use of HMC to those who understand the theory and have the programming skills to implement the algorithm. But it does not have to be so.

The emergence of modern Bayesian software such as \proglang{Stan} {\citep{carpenter_stan:_2017}} has, to some extent, alleviated these difficulties. \proglang{Stan} is a powerful and versatile programming language that has a syntax similar to that of \proglang{WinBUGS}, but uses HMC instead of Gibbs sampling to generate  posterior samples \citep{gelmanstan2015}. \proglang{Stan} translates its code to a lower-level language to maximize speed and efficiency. Importantly, it automates the tuning of HMC parameters and thus significantly reduces the burden of implementation. For \proglang{R} and \proglang{Python} users, packages have been created that allow \proglang{Stan} be called from those languages. For people who are familiar with \proglang{WinBUGS} and comfortable with programming in probabilistic terms, \proglang{Stan} is an ideal choice for HMC implementation. But for beginners who want to learn HMC, \proglang{Stan} can come across as a ``black box''. Other high-performance software, such as \pkg{PyMC} and \pkg{Edward} \citep{salvatierprobabilistic2016,tranedward2016}, presents similar challenges. While scalability and efficiency are often the foremost considerations in software development, a good understanding of the methodology is more essential to learners, as it instills confidence in the practical use of new methods. 

The objectives of the current paper are largely pedagogical, i.e., helping practitioners learn HMC and its algorithmic ingredients. Toward that end, we developed a general-purpose \proglang{R} function \texttt{hmc} for the fitting of common statistical models. We also present details of HMC parameter tuning for those who are interested in writing and implementing their own programs. Multiple examples are presented, with accompanying \proglang{R} code. We have assembled all of the learning material, including the necessary HMC functions, example code, and data in an \proglang{R} package, \pkg{hmclearn}, for convenience of the readers.

\section{Markov Chain Monte Carlo: The Basics}

MCMC is a broad class of computational tools for integral approximation and posterior sample generation. In Bayesian analysis, MCMC algorithms are primarily used to simulate samples for approximation of the posterior distribution. 

In Bayesian analysis, estimation and inference of the parameter of interest are made based on the observed data $\mathcal{D}$ together with the \textit{a priori} information that one has on the parameters of interest  $\boldsymbol\theta = (\theta_1, ..., \theta_k)^T\in \mathbb{R}^k$.  The posterior distribution $f(\boldsymbol\theta | \mathcal{D})$ combines both the data and prior information in accordance to the Bayes formula,  and is proportional to the product of the likelihood function $f(\mathcal{D} | \boldsymbol\theta)$ and the prior density $f(\boldsymbol\theta)$ \citep{carlinbayesian2008} , 
$$
\begin{aligned}
f(\boldsymbol\theta | \mathcal{D}) &= \frac{f(\mathcal{D}|\boldsymbol\theta)f(\boldsymbol\theta)}{\int f(\mathcal{D}|\boldsymbol\theta)f(\boldsymbol\theta)d\boldsymbol\theta}, \\
&\propto f(\mathcal{D}|\boldsymbol\theta) f(\boldsymbol\theta).
\end{aligned}
$$

The integral in the denominator is usually difficult to evaluate.  But since the denominator is constant with respect to $\boldsymbol\theta$, one could work with the unnormalized posterior $f(\mathcal{D} | \boldsymbol\theta)f(\boldsymbol\theta)$. In the absence of an explicit expression of the posterior, approximating it with simulated samples following $f(\boldsymbol\theta | \mathcal{D})$ becomes a desirable alternative. 

\subsection{Metropolis-Hastings}

Metropolis algorithm is the first widely-used MCMC method for generating Markov Chain samples following $f(\boldsymbol\theta |\mathcal{D} )$. The method originated from a physics application in the 1950's \citep{metropolisequation1953}, and was further extended nearly two decades later by \cite{hastingsmonte1970}, thus giving rise to the name of Metropolis-Hastings (MH) algorithm.  We begin with a brief description of MH, as HMC was built on a similar concept.  

MH generates a sequence of values of $\boldsymbol\theta$ that form a Markov chain, whose values can be used to approximate a posterior density $f(\boldsymbol\theta | \mathcal{D})$.  For brevity, we drop $\mathcal{D}$ from the expression and write the posterior simply as $f(\boldsymbol\theta)$. Values in the Markov chain $\boldsymbol\theta^{(t)}$ are indexed by $t = 0,1, .., N$, where $\boldsymbol\theta^{(0)}$ is a user or program-specified starting value.  

MH defines a transition probability that assures the Markov chain is \emph{ergodic} and satisfies \emph{detailed balance} and \emph{reversibility} \citep{chibunderstanding1995}. These technical conditions are put in place to ensure the chain samples from the full support of $\boldsymbol\theta$ without bias.  

In MH, values of $\boldsymbol\theta^{(t)}$ in the chain are defined in part by a proposal density $q(\boldsymbol\theta^{\text{PROP}} | \boldsymbol\theta^{t-1})$, where $\boldsymbol\theta^{\text{PROP}}$ is a proposal for the next value in the chain.  This proposal density is conditioned on the previous value $\boldsymbol\theta^{(t-1)}$.  A variety of proposal functions can be used, with random walk proposals being the most common choice.  

\begin{algorithm}
\caption{Metropolis-Hastings}\label{Metropolis-Hastings}
\begin{algorithmic}[1]
\Procedure{MH}{$\boldsymbol\theta^{(0)}, f(\boldsymbol\theta), q(\boldsymbol\theta^{(1)}|\boldsymbol\theta^{(2)}), N$} 
   \State Calculate $f(\boldsymbol\theta^{(0)})$ 
   \For{$t = 1, ..., N$} 
      \State $\boldsymbol\theta^{\text{PROP}} \gets q(\boldsymbol\theta^{\text{PROP}} | \boldsymbol\theta^{(t-1)})$ 
      \State $u \gets U(0, 1)$
      \State $\alpha = \min\left(1, \frac{f(\boldsymbol\theta^{\text{PROP}})q(\boldsymbol\theta^{(t-1)}|\boldsymbol\theta^{\text{PROP}})}{f(\boldsymbol\theta^{(t-1)})q(\boldsymbol\theta^{\text{PROP}}|\boldsymbol\theta^{(t-1)})} \right)$ 
      \State If $\alpha < u$, then $\boldsymbol\theta^{(t)} \gets \boldsymbol\theta^{\text{PROP}}$.  Otherwise, $\boldsymbol\theta^{(t)} \gets \boldsymbol\theta^{(t-1)}$ 
   \EndFor\label{markovendfor}
   \State \textbf{return} $\boldsymbol\theta^{(1)} ... \boldsymbol\theta^{(N)}$ 
\EndProcedure
\end{algorithmic}
\end{algorithm}

In MH, a proposal is accepted with probability
\begin{equation}
\alpha = \min\left(1, \frac{f(\boldsymbol\theta^{\text{PROP}})q(\boldsymbol\theta^{(t-1)}|\boldsymbol\theta^{\text{PROP}})}{f(\boldsymbol\theta^{(t-1)})q(\boldsymbol\theta^{\text{PROP}}|\boldsymbol\theta^{(t-1)})} \right),\label{alpha}
\end{equation}

When $q$ is symmetric i.e., $q(\boldsymbol\theta^{(t-1)}|\boldsymbol\theta^{\text{PROP}}) = q(\boldsymbol\theta^{\text{PROP}}|\boldsymbol\theta^{(t-1)})$, this simplifies to  
$$
\alpha = \min\left(1, \frac{f(\boldsymbol\theta^{\text{PROP}})}{f(\boldsymbol\theta^{(t-1)})} \right),
$$
which is used in the original Metropolis algorithm.

The denominator in the posterior is constant with respect to $\boldsymbol\theta$.  As such, the ratio of posterior densities at two different points $\boldsymbol\theta^{PROP}$ and $\boldsymbol\theta^{(t-1)}$ can be compared even when the denominator is unknown, with the denominators being cancelled out.  Because a derivation of the full posterior distribution (numerator and denominator) is not necessary to implement MH (and HMC, as we will see),  data analysts have considerable flexibility to select models of their liking. 

The acceptance rate $\alpha$ in (\ref{alpha}) is an important gauge of the efficiency of an MH algorithm. A careful examination of $\alpha$'s roles gives a more intuitive understanding of the algorithm:
\begin{enumerate}
\item When $f(\boldsymbol\theta^{PROP}) \geq f(\boldsymbol\theta^{(t-1)})$, the proposal  $f(\boldsymbol\theta^{PROP})$ represents a ``more likely'' value  than  the previous value $\boldsymbol\theta^{(t-1)}$, as quantified by the density functions.  When this occurs, the proposal is always accepted (i.e. with probability 1).  
\item When $f(\boldsymbol\theta^{PROP}) < f(\boldsymbol\theta^{(t-1)})$, the proposal $\boldsymbol\theta^{PROP}$ has a lower density in comparison to the previous value, and we accept the proposal at random with probability $\alpha\in(0,1)$, which indicates the relative likelihood of observing $\boldsymbol\theta^{PROP}$ from $f$, as compared to $\boldsymbol\theta^{(t-1)}$. The larger the $\alpha$, the greater the chance of accepting $\boldsymbol\theta^{PROP}$. If the proposal is not accepted, the proposal will be discarded and the chain will remain in place $\boldsymbol\theta^{(t)} := \boldsymbol\theta^{(t-1)}$, and we will start with a new proposal.  
\end{enumerate}

With such a scheme, the algorithm frequents regions of \emph{higher} posterior density, while occasionally visiting the low density areas (e.g., tails in one-dimensional situations). Provided the algorithm satisfies the conditions for ergodicity \citep{tierney_markov_1994} and runs a sufficient number of iterations, the empirical distribution of the MCMC chain samples should approximate the true posterior density.  The simulated values can therefore be used for estimation and inference based on the posterior distribution.  See \cite{carlinbayesian2008} \cite{chibunderstanding1995} \cite{gelmanbayesian2013} for additional details on MH.

\subsection{Limitations of Metropolis-Hastings}

The theoretical requirements for using MH are quite minimal, thus making it an attractive choice for Bayesian inference.  Limitations of MH are primarily computational.  With randomly generated proposals, it often takes a large number of iterations to get into areas of higher posterior density. Even efficient MH algorithms sometimes accept less than $25\%$ of the proposals \citep{robertsweak1997}.  In lower dimensional situations, increased computational power may compensate the lower efficiency to some extent. But in higher dimensional and more complex modeling situations, bigger and faster computers alone are rarely sufficient to overcome the challenge.  

Gibbs sampling can be a viable and more efficient alternative to MH in some situations \citep{gemanstochastic1984}. In fact, several popular software platforms, such as   WinBUGS and JAGS, use Gibbs to generate posterior samples  \citep{lunnwinbugs2000,plummerjags2003}. Gibbs' requirement for explicitly expressed conditional posterior densities, however, has prevented it from being used in many practical situations. In addition to this restriction, Gibbs also has its own efficiency limitations \citep{robertbayesian2001}.  It is in this context that HMC emerges as a preferred alternative for Bayesian analysis.

\section{Hamiltonian Monte Carlo}

Hamiltonian Monte Carlo improves the efficiency of MH by employing a guided proposal generation scheme. More specifically, HMC uses the gradient of the log posterior to direct the Markov chain towards regions of higher posterior density, where most samples are taken.  As a result, a well-tuned HMC chain will accept proposals at a much higher rate than the traditional MH algorithm \citep{robertsweak1997}. 

It is important to note that although the HMC algorithm frequently samples in regions of higher density, referred to as the \emph{typical set} \citep{betancourtconceptual2017}, it still samples the tail areas properly.  While both MH and HMC produce ergodic Markov chains, the mathematics of HMC is substantially more complex than that of MH.  In this paper, we provide a less technical introduction of the ideas behind HMC. More technical expositions can be found elsewhere \citep{nealhandbook2011,betancourtconceptual2017}. 

\subsection{The idea}

\begin{figure}
\begin{center}
\includegraphics[width=0.9\textwidth]{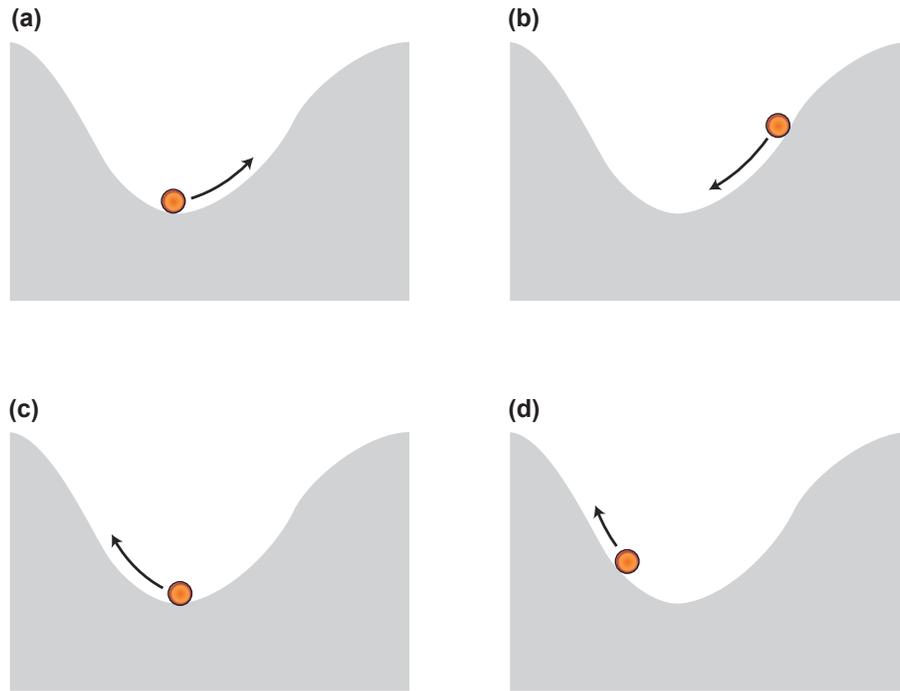}
\end{center}
\caption{\small{One-dimensional HMC example - movement of an object on a smooth, frictionless curve.  (a) We apply a force with randomly generated direction and strength to the object. This object acquires a certain amount of kinetic energy, which makes it move in the direction of the applied force.  The momentum, proportional to the object’s velocity, changes throughout the path of the curve. When the object moves up along the curve, the velocity of the object and its momentum decrease.  Its kinetic energy converts to potential energy, while the total energy remains constant.  (b)  The object will stop at a point when all of its kinetic energy is converted to potential energy.  The potential energy then makes the object move in the opposite direction, converting its potential energy back to kinetic energy.  (c) At the lowest point of the curve, all of the energy is in the kinetic form (peak velocity/momentum), which pushes the object up to the left side of the curve.  (d) As the object goes up on the curve, its kinetic energy again converts to potential energy, until all is in the form of potential energy. Then, the object would stop and then slide back as guided by its potential energy.  Since the surface is frictionless, the total energy remains constant throughout these repeated movements.}}
\label{fig:idea}
\end{figure}

The methods one uses to generate proposals strongly influences the efficiency of MCMC. Suppose $f(\theta)$ is a one-dimensional posterior density function,  and $-\log f(\theta)$ assumes the shape of an inverse bell-shaped curve as depicted by Figure~\ref{fig:idea}. To generate $\theta$  in a region of high posterior density, one needs to sample $\theta$ in the region corresponding to the lower values of  $-\log f(\theta)$; the region can be reached with the guidance of the gradient of $-\log f(\theta)$. In a sense, the approach is analogous to the movement of a hypothetical object on a frictionless curve, where the object traverses and lingers at the bottom of the valley while occasionally visiting the higher grounds on both sides. In classical mechanics, such movements are described by the Hamiltonian equations, where the exchanges of kinetic and potential energy dictate the object's location at any given moment.  


In a Hamiltonian system, the horizontal and vertical positions are given by $(\theta,p)$. In MCMC,  we are interested in $\theta$. The parameter $p$, which is often referred to as the \emph{momentum}, is an auxiliary quantity that we use to simulate $\theta$ under the Hamiltonian equations.  

\subsection{The Hamiltonian Equations}

We introduce HMC in a generic MCMC setting, where $\boldsymbol\theta$ follows the posterior density $f(\boldsymbol\theta)$ of interest, and the momentum $\bold{p}$ is generated from a parametric distribution. The momentum matches the dimensionality of $\boldsymbol\theta$ as a vector of length $k$.

We write the Hamiltonian function as $H(\boldsymbol\theta, \bold{p})$,  which consists of \emph{potential} energy $U(\boldsymbol\theta)$ and \emph{kinetic} energy $K(\bold{p})$:  $H(\boldsymbol\theta, \bold{p}) =  U(\boldsymbol\theta) + K(\bold{p})$, where $\bold{p}$ and $\boldsymbol\theta \in \mathbb{R}^k$. 

In statistical applications of MCMC, we are primarily interested in generating $\boldsymbol\theta$ from a given distribution $f(\boldsymbol\theta)$. To do so, we let $U(\boldsymbol\theta):= - \log f(\boldsymbol\theta)$. Such a designation would ensure $\boldsymbol\theta$ generated from the Hamiltonian function follows the desired distribution. For momentum,  we typically assume  $\bold{p} \sim N_k(0, \bold{M})$, where $\bold{M}$ is a user-specified covariance matrix. 

Under this formulation, we have

\begin{equation}
H(\boldsymbol\theta, \bold{p}) = - \log f(\boldsymbol\theta) + \frac{1}{2}\bold{p}^T \bold{M}^{-1} \bold{p}.\label{eq:Hamiltonian}
\end{equation}

Over time, HMC travels on trajectories that are governed by the following first-order differential equations, known as the \emph{Hamiltonian equations}

\begin{equation}
\begin{split}
\frac{d\bold{p}}{dt} &= -\frac{\partial H(\boldsymbol\theta, \bold{p})}{\partial\boldsymbol\theta}= -\frac{\partial U(\boldsymbol\theta)}{\partial\boldsymbol\theta} = \nabla_{\boldsymbol\theta} \log f(\boldsymbol\theta), \\
\frac{d\boldsymbol\theta}{dt} &= \frac{\partial H(\boldsymbol\theta, \bold{p})}{\partial\bold{p}} = \frac{\partial K(\bold{p})}{\partial\bold{p}} = \bold{M}^{-1}\bold{p},
\end{split}
\label{eq:diffeq}
\end{equation}

where  $\nabla_{\boldsymbol{\theta}} \log f(\boldsymbol{\theta})$ is the gradient of the log posterior density. A solution to the Hamiltonian equations is a function that defines the path of $(\boldsymbol\theta,\bold{p})$ from which specific values of $\boldsymbol\theta$ could be sampled. Within an MCMC iteration, we sample a value $\boldsymbol\theta$ from this path.  The randomness of the MCMC samples comes from the momentum $\bold{p}\sim N_k(0, \bold{M})$ and the specific $\boldsymbol\theta$ value we choose.

\subsection{Solving the Hamiltonian Differential Equations}

Solving the Hamiltonian equations, therefore, becomes a critical step in HMC simulation.   A standard approach for solving differential equations is Euler's method, which produces a discrete function that approximates the solution at each time \textit{t}.  Values of $(\boldsymbol\theta, \bold{p})$ that satisfy the Hamiltonian equations would be legitimate values for the HMC.  But as  \citet{nealhandbook2011}  have noted, errors tend to accumulate in Euler's method, especially after a larger number of steps. In HMC, one often has to take a larger number of steps to ensure the new proposal is sufficiently far from the location of the previous sample. 

The \emph{leapfrog} method is a good alternative to the standard Euler's method  for approximating the solutions to Hamiltonian equations \citep{ruth1983canonical}. The leapfrog algorithm modifies Euler's method by using a discrete step size $\epsilon$ individually for $\bold{p}$ and $\boldsymbol\theta$, with a full step $\epsilon$ in $\boldsymbol\theta$ sandwiched between two half-steps $\epsilon/2$ for $\bold{p}$,

\begin{equation}
\begin{split}
\bold{p}(t + \epsilon/2) &= \bold{p}(t) + (\epsilon/2)\nabla_{\boldsymbol\theta}\log f(\boldsymbol\theta(t)), \\
\boldsymbol\theta(t + \epsilon) &= \boldsymbol\theta(t) + \epsilon \bold{M}^{-1}\bold{p}(t + \epsilon/2), \\
\bold{p}(t + \epsilon) &= \bold{p}(t + \epsilon/2) + (\epsilon/2)\nabla_{\boldsymbol\theta} \log f(\boldsymbol\theta(t + \epsilon)).
\end{split}\label{eq:leapfrog}
\end{equation}

For HMC, multiple leapfrog steps are typically required to move a sufficient distance to the next proposal.  Research has shown that discrete approximations remain accurate, even after many steps. The stability of the leapfrog algorithm is due to the leapfrog's symplectic property. \citep{channell_symplectic_1990,betancourtconceptual2017}. Symplecticity ensures that the volume of the support is preserved when mapping from one point to another, such as through one or more consecutive iterations of the leapfrog algorithm\citep{nealhandbook2011}. 

For a given momentum vector $\bold{p}$ within an HMC iteration, the path defined by the Hamiltonian equations is deterministic.  Proposals generated from an exact solution of these equations, if achievable, would always be accepted.  But since our solution from the leapfrog is an approximation, a Metropolis style accept/reject step is added to ensure the newly generated proposal does not deviate too far from the specified Hamiltonian $H(\boldsymbol\theta, \bold{p})$. The acceptance rate of HMC proposals is therefore less than 100\%, but generally higher than that of the Metropolis algorithm.

\subsection{Hamiltonian Monte Carlo Algorithm}

\begin{figure}
\includegraphics[width=0.9\textwidth]{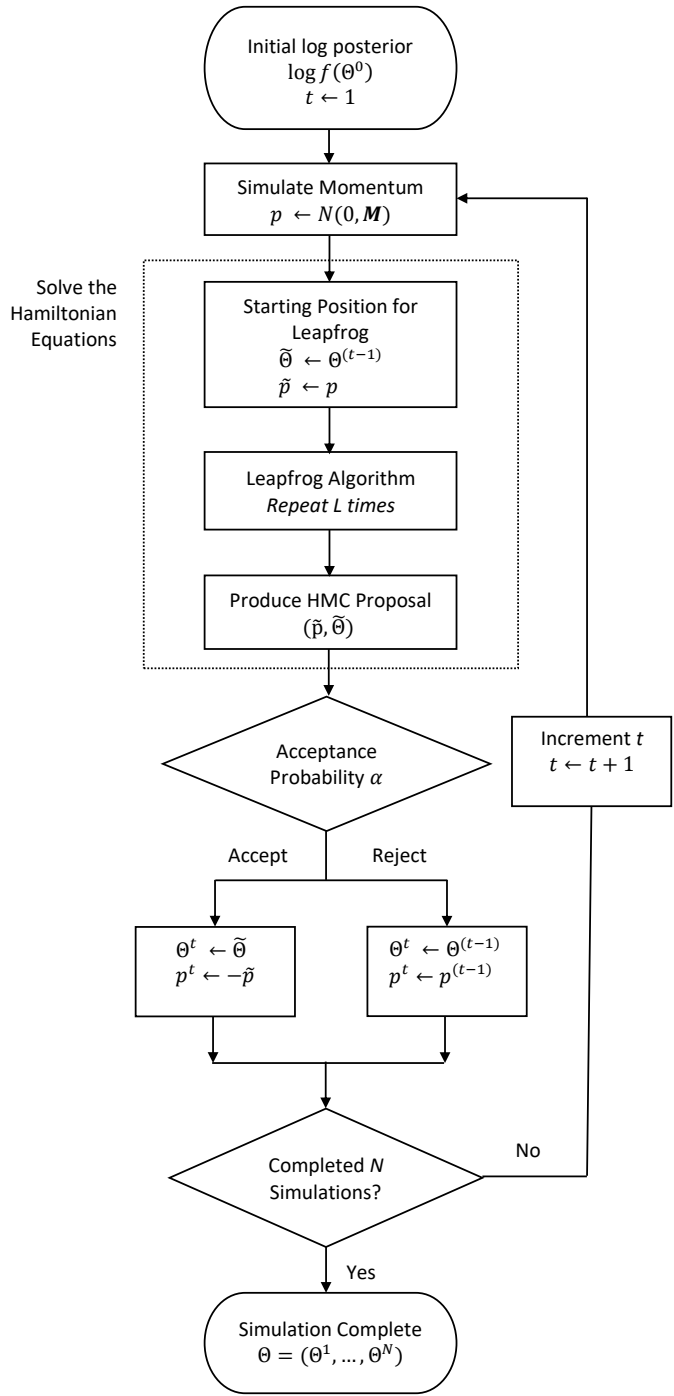}
\caption{Main Steps of the Hamiltonian Monte Carlo Method}
\label{fig:HMCflow}
\end{figure}

The flowchart in Figure \ref{fig:HMCflow} shows the key steps in HMC.  Initial values for $\boldsymbol\theta$ and $\bold{p}$ are required to start the algorithm.  
With $\boldsymbol\theta^{(0)}$ and $\bold{p}^{(0)}$ specified, the leapfrog algorithm is used to find approximate solutions to the Hamiltonian equations.  The leapfrog solutions define the path of $(\boldsymbol\theta, \bold{p})$ over time within an iteration.

Typically, multiple steps, each of length $\epsilon$, are taken to generate an HMC proposal.  Parameter $L$ represents the number of steps.  While $L$ is often fixed to a positive integer value, some randomness can be introduced to ensure a valid exploration of the space of $(\boldsymbol\theta, \bold{p})$.  A generic HMC is given below in Algorithm \ref{HMC}.

\begin{algorithm}
\caption{Hamiltonian Monte Carlo}\label{HMC}
\begin{algorithmic}[1]
\Procedure{HMC}{$\boldsymbol\theta^{(0)}, \log f(\boldsymbol\theta), \bold{M}, N, \epsilon, L$} 
   \State Calculate $\log f(\boldsymbol\theta^{(0)})$
   \For{$t = 1, ..., N$}
      \State $ {\bold{p}}  \gets N(0, \bold{M})$
      \State $\boldsymbol\theta^{(t)} \gets \boldsymbol\theta^{(t-1)}, \tilde{\boldsymbol\theta} \gets \boldsymbol\theta^{(t-1)}, \tilde{\bold{p}} \gets {\bold{p}  }$ 
      \For{$i = 1, ..., L$}
        \State $\tilde{\boldsymbol\theta}, \tilde{\bold{p}} \gets \text{Leapfrog}(\tilde{\boldsymbol\theta}, \tilde{\bold{p}}, \epsilon, \bold{M})$
      \EndFor\label{leapfrogfor}

      \State $\alpha \gets \min{\left(1, \frac{\exp(\log{f(\tilde{\boldsymbol\theta}}) - \frac{1}{2}{\bold{\tilde{p}}^T \bold{M}^{-1} {\bold{\tilde{p}})}}}{\exp(\log f(\tilde{\boldsymbol\theta}^{(t-1)}) - \frac{1}{2} {\bold{p}^T} \bold{M}^{-1} \bold{p})} \right)}$ 

      \State With probability $\alpha$, $\boldsymbol\theta^{(t)}\gets \tilde{\boldsymbol\theta}$ and $\bold{p}^{(t)} \gets -\tilde{\bold{p}}$ 
   \EndFor\label{ehmcendfor}
   \State \textbf{return} $\boldsymbol\theta^{(1)}, ..., \boldsymbol\theta^{(N)}$
\Function{Leapfrog}{$\boldsymbol\theta^*, \bold{p}^*, \epsilon, \bold{M}$}
\State $\tilde{\bold{p}} \gets \bold{p}^* + (\epsilon/2)\nabla_{\boldsymbol\theta}\log f(\boldsymbol\theta^*)$
\State $\tilde{\boldsymbol\theta} \gets \boldsymbol\theta^* + \epsilon\bold{M}^{-1}\tilde{\bold{p}}$
\State $\tilde{\bold{p}} \gets \tilde{\bold{p}} + (\epsilon/2)\nabla_{\boldsymbol\theta}\log f(\tilde{\boldsymbol\theta} )$
\State \textbf{return} $ \tilde{\boldsymbol\theta}, \tilde{\bold{p}}$
\EndFunction
\EndProcedure
\end{algorithmic}
\end{algorithm}

As with other valid MCMC algorithms, HMC's transition probability is designed to meet the theoretical requirements for detailed balance and reversibility. These conditions ensure that our HMC samples provide a valid representation of the posterior distribution. If we denote the transition probability from $\boldsymbol\theta^{(t)}$ to $\boldsymbol\theta^{(t+1)}$ as $T(\boldsymbol\theta^{(t)}, \boldsymbol\theta^{(t+1)})$, then detailed balance requires that $f(\boldsymbol\theta^{(t)}) T(\boldsymbol\theta^{(t)}, \boldsymbol\theta^{(t+1)}) = f(\boldsymbol\theta^{(t+1)}) T(\boldsymbol\theta^{(t+1)}, \boldsymbol\theta^{(t)})$. The HMC transition probability includes two components to ensure that detailed balance and reversibility hold true:
 
 \begin{enumerate}
 \item the accept/reject step, and 
 \item the negation of the momentum after the final leapfrog step.  
 \end{enumerate}

The negated momentum illustrates the reversibility of HMC transitions, which can be demonstrated by stepping through the leapfrog from the proposed state to the original state. \cite{tierney_markov_1994} described the theoretical requirements for MCMC algorithms in general, while \cite{betancourtconceptual2017} provided a detailed exposition specific to HMC.

In Section~\ref{package}, we describe a general-purpose function \texttt{hmc}  in our proposed package. Within the package, the gradient functions for commonly used generalized linear mixed effect models under the default priors are provided. The \texttt{hmc} function can also take user-defined posterior density and gradient functions for non-standard statistical models. In situations where analytical derivation of gradient functions is infeasible, one could consider using numerical auto-differencing functions. Automated differencing libraries capable of calculating the gradient exactly such as the \proglang{Stan} math library \citep{carpenter_stan_2015}, also called Autodiff, are appropriate for direct use in HMC applications.

\hypertarget{tuning}{%
\subsection{HMC Tuning for Improved Efficiency}\label{tuning}}

The efficiency of an HMC algorithm can be improved through parameter tuning and reparameterization. HMC tuning involves selection and adjustment of the various HMC parameters. Two parameters that need to be specified are the step size $\epsilon$ and the number of leapfrog steps $L$.  Elements in the covariance matrix $\bold{M}$ may also be adjusted from the default identity matrix for efficiency improvement.  

It is generally a good practice to set \(\epsilon\) to a smaller value relative to the magnitude of the parameter of interest.  A smaller \(\epsilon\) results in closer approximations and thus higher acceptance rates. But a small  $\epsilon$ must be coupled with a large $L$ to ensure the trajectory length $\epsilon L$ is large enough to move the simulated parameter to a distant point in the distribution. On the other hand, if $\epsilon L$ is too large the trajectory is likely to circle back, causing waste in simulation. To tune $\epsilon$ and $L$ is to find the right combinations of these values, which are usually chosen via monitoring the acceptance rate. \citet{nealhandbook2011} suggested an optimal acceptance rate is approximately 65\%.  
At the same time, it is often helpful to examine the trace plots of the MCMC samples for signs of autocorrelation. Slow-moving chains with stronger autocorrelation often indicate insufficient $\epsilon L$. While \(\epsilon\) and \(L\) can be tuned jointly, most analysts choose to select the step size first, then under a given step size, they fine-tune the number of steps per leapfrog \(L\).

Additional adjustments may be made to the tuning parameters beyond these basic steps.  For example,  one could use different values of $\epsilon$ for each of the $k$ parameters in $\boldsymbol\theta$ to increase the sampling efficiency.  The \texttt{hmc} function in \pkg{hmclearn} allows setting $\epsilon$ to a vector instead of a single number to give analysts the flexibility to use different step sizes for different parameters. The parameter for the number of steps $L$ must be a natural number. However, randomly chosen $L$ could be used to guard against periodicity of the Markov chain.  The step size $\epsilon$ may also be randomized.  In the \texttt{hmc} function, random $\epsilon$ and $L$ can be automatically applied via parameter setting. A useful algorithm known as the No U-Turn Sampler (NUTS)  automatically selects $L$ for each sample; NUTS is a commonly used alternative to manual parameter tuning \citep{hoffman_no-u-turn_2014}.  

The efficiency of sampling in the standard HMC algorithm can also be improved for multivariate models when the parameters have an orthogonal basis. One common method of ensuring an orthogonal basis involves applying QR decomposition \citep{voss_introduction_2013} .  In many statistical models, especially linear models, the design matrix helps to define the model itself and is central to the model fitting computation. In HMC, QR decomposition is often applied to the design matrix to create the orthogonal basis for sampling. Applying this transformation in practice can improve the computational efficiency of HMC for many models \citep{stan_development_team_stan_2017}.   After the simulation is complete, the MCMC samples are transformed back to the original basis for inference.

\section{A Package for Learning HMC}\label{package}

HMC presents considerable challenges to beginners attempting to learn the algorithm.  First, the method can be difficult to comprehend because its idea originated from physics applications of the Hamiltonian equations. Second, it is often difficult to learn the inner working of HMC from programs such as \proglang{Stan}, because they are not designed as teaching tools. In fact,  \proglang{Stan} specifies models in a probabilistic syntax and shields users from the actual HMC steps. 

In this paper, we present an \proglang{R} package \pkg{hmclearn} to provide users with the software tools to \emph{learn} the intricacies of the HMC, through explicit specification of log posterior and gradient functions, as well as parameter tuning. It is designed to give user a hands-on experience for implementing HMC analysis for a broad class of statistical models.  Once users have understood and mastered the essential HMC steps, they could go on to write their own code for  specific applications. To download \pkg{hmclearn}, go to \url{https://cran.r-project.org/web/packages/hmclearn/index.html}.

The core function in \pkg{hmclearn} is  \texttt{hmc}, which is a general-purpose function for MCMC sample generation by using the HMC method. This function takes user-defined log posterior and gradient functions as inputs and produces MCMC samples. Here we do not ask for an explicit specification of prior $f(\boldsymbol{\theta})$ as an input function. Instead, we let users define their log posterior $\log f(\boldsymbol{\theta|}\boldsymbol{y})= \log f(\boldsymbol{y}|\boldsymbol{\theta}) + \log f(\boldsymbol{\theta})$, which includes $f(\boldsymbol{\theta})$. Such a design reduces the number of required input functions, while preserving users' flexibility in choosing different priors.  

Other input parameters to \texttt{hmc} include the number of samples $N$, the step size $\epsilon$,  the number of leapfrog steps $L$, and the Mass matrix $\bold{M}$.  These are the essential elements to start an HMC simulation, but the user will typically need to adjust at least some of these parameters to tailor the simulation to their specific applications.  Users are required to provide their own starting values for $\boldsymbol\theta$ when using the $\texttt{hmc}$ function for their own applications. Examples of log posterior and gradient functions are provided in \pkg{hmclearn} for various generalized linear mixed effect models, which can be used as templates for less standard models.  

Running multiple MCMC chains is often desirable to determine if each chain converges to the same distribution of $\boldsymbol\theta$.  Since modern computers almost universally have multiple core processors, parallel processing can be an efficient way to run multiple chains at the same time.  To that end, \pkg{hmclearn} includes parameters to enable parallel processing as well as multiple chains.  

Finally, a variety of Bayesian graphical functions are provided based on the \texttt{bayesplot} package \citep{gabry2016bayesplot}.  Functionalities incorporated in \pkg{hmclearn} include trace plots, histograms, density plots, and credible interval plots.  The integrated functions comprise the core diagnostic plotting functions typical for MCMC applications.  Additional diagnostics can be programmed directly or called based on the output of the \texttt{hmc} function.  

\section{HMC in Statistical Models}

\subsection{A general process}

\begin{figure}
\begin{center}
\includegraphics[width=0.9\textwidth]{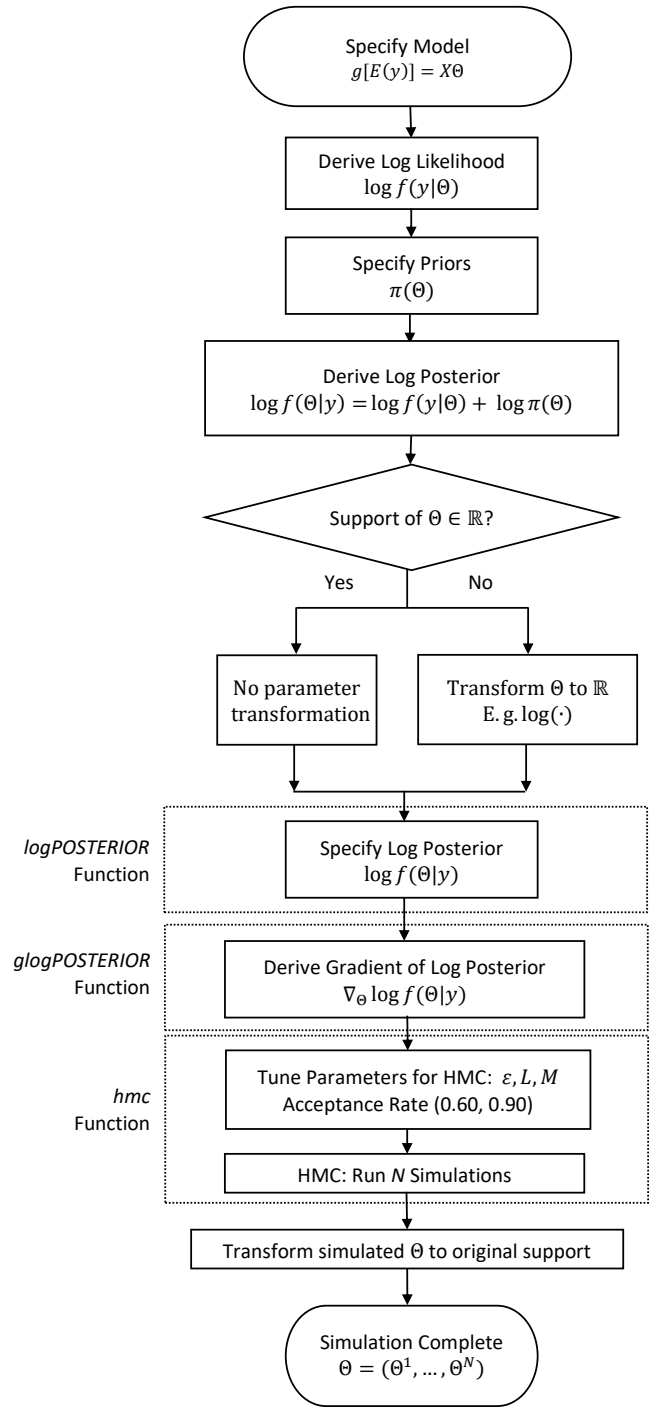}
\end{center}
\caption{{Major steps of HMC implementation}}
\label{fig:steps}
\end{figure}

In this section, we discuss the general steps of HMC implementation in statistical models. We describe the process through examples of generalized linear models. The major steps required to fit a statistical model are summarized in Figure~\ref{fig:steps}. Following the steps illustrated in the diagram, one could generate HMC samples with user-specified posterior and gradient functions, by using the  \texttt{hmc} function in the \pkg{hmclearn} package. 


\hypertarget{hmc-examples}{%
\subsection{Examples}\label{hmc-examples}}

We present three examples to illustrate how to fit various linear models using HMC. Our notation for these examples reflects the programming of the sample log posterior and gradient functions in \pkg{hmclearn}. This programming uses matrix and vector multiplication instead of for loops, which can be computationally slow in R.  

\hypertarget{linear-regression}{%
\subsubsection{Example 1: Linear Regression}\label{linear-regression}}

We consider a linear regression model $y_i = \bold{x}_i^T\boldsymbol\beta + \epsilon_i$, where $y_i$ is the response for the \textit{i}th subject, $i = 1, ..., n$, and $\bold{y} = (y_1, ..., y_n)^T$ is a vector of responses. The covariate values for the \textit{i}th subject are $\bold{x}_i^T = (x_{i0}, ..., x_{iq})$, where $x_{i0}$ is frequently set to one as an intercept term for all subjects. We write the full design matrix as $\bold{X} = (\bold{x}_1^T, ..., \bold{x}_n^T)^T \in \mathbb{R}^{n\times(q+1)}$. The regression coefficients for the $q$ covariates plus an intercept are written as $\boldsymbol\beta = (\beta_0, ..., \beta_q)^T$.  The error term for each subject is $\epsilon_i$. All error terms $\boldsymbol\epsilon = (\epsilon_1, ..., \epsilon_n)^T$ are assumed to be independent and normally distributed with mean zero and constant variance $\sigma_\epsilon^2$.

The log likelihood for linear regression, omitting the constants, can be written as
$$
\log f(\bold{y} | \boldsymbol\beta, \sigma_\epsilon^2)\propto-n\log\sigma_\epsilon - \frac{1}{2\sigma_\epsilon^2} (\mathbf{y} - \mathbf{X}\boldsymbol\beta)^T(\mathbf{y} - \mathbf{X}\boldsymbol\beta).
$$

We specify a multivariate normal prior for $\boldsymbol\beta$ with covariance matrix $\sigma_\beta^2 \bold{I}$ where $\sigma_\beta^2$ is a hyperparameter set by the analyst, and an inverse gamma (IG) prior for $\sigma_\epsilon^2$.   The IG prior has hyperparameters $a$ and $b$, which are also set by the analyst. We write
$$
\begin{aligned}
f(\boldsymbol\beta | \sigma_\beta^2) &\propto \exp\left(-\frac{\boldsymbol\beta^T  \boldsymbol\beta}{2\sigma_\beta^2} \right) \quad\textrm{ and}\quad
f(\sigma_\epsilon^2 | a, b) &= \frac{b^a}{\Gamma{(a)}} (\sigma_\epsilon^2)^{-a-1}\exp\left(-\frac{b}{\sigma_\epsilon^2}\right).
\end{aligned}
$$

The support of $\sigma_\epsilon^2$ is $(0, \infty)$. We apply a logarithmic transformation to expand the support to $\mathbb{R}$. We have
$$
\begin{aligned}
\gamma &= \log \sigma_\epsilon^2, \quad
\sigma_\epsilon^2 = g^{-1}(\gamma) = e^\gamma, \\
f(\gamma | a, b) 
&= \frac{b^a}{\Gamma(a)}\exp\left(-a\gamma-\frac{b}{e^\gamma} \right), \\
\log f(\gamma | a, b) &\propto -a\gamma - be^{-\gamma}.
\end{aligned}
$$

The log posterior is proportional to the log likelihood plus the log prior, 
$$
\begin{aligned}
\log f(\boldsymbol\beta, \gamma | \bold{y}, \bold{X}, \sigma_\beta^2, a, b)
&\propto -\left (\frac{n}{2} + a  \right)\gamma - \frac{e^{-\gamma}}{2} (\mathbf{y} - \mathbf{X}\boldsymbol\beta)^T(\mathbf{y} - \mathbf{X}\boldsymbol\beta)-\frac{\boldsymbol\beta^T \boldsymbol\beta}{2\sigma_\beta^2} - b e^{-\gamma}.
\end{aligned}
$$

The parameters of interest are defined as $\boldsymbol\theta := (\beta_0, ..., \beta_{q}, \gamma)^T$, where $k = q + 2$.  To fit this model using \texttt{hmc}, the user must provide a function for the log posterior where the first function parameter is a vector for the parameters of interest $\boldsymbol\theta$. Additional function parameters can be included for the data and hyperparameters. An example log posterior function for this model and specification of priors is included in \pkg{hmclearn}. 

The Hamiltonian function (\ref{eq:Hamiltonian}) is composed of the log posterior and the log density function of the momentum, where $\bold{p} \sim N_k(0, \bold{M})$. Writing the Hamiltonian function for our linear regression model is straightforward once the log posterior is developed,  
$$
\begin{aligned}
H(\boldsymbol\theta, \bold{p}) &= H(\boldsymbol\beta, \gamma, \bold{p}) \propto \log f(\boldsymbol\beta, \gamma | \bold{y}, \bold{X}, \sigma_\beta^2, a, b)+ \frac{1}{2}\bold{p}^T \bold{M}^{-1} \bold{p}.
\end{aligned}
$$
With the Hamiltonian function explicitly defined, we can write the Hamiltonian equations (\ref{eq:diffeq}) for this particular model. 

The steps of the leapfrog algorithm are integrated with \texttt{hmc} in a self-contained function. This function requires, as an input, a separate standalone function that returns a vector for the gradient of the log posterior. As with the log posterior function, the first function parameter must be a vector for $\boldsymbol\theta$. The gradient functions for the model in this example are also included in \pkg{hmclearn},  
$$
\begin{aligned}
\nabla_{\boldsymbol\beta} \log f(\boldsymbol\beta, \gamma | \bold{y}, \bold{X}, \sigma_\beta^2, a, b) &\propto e^{-\gamma} \mathbf{X}^T ( \mathbf{y} - \mathbf{X}\boldsymbol\beta)- \boldsymbol\beta / \sigma_\beta^2  , \\
\nabla_\gamma \log f(\boldsymbol\beta, \gamma | \bold{y}, \bold{X}, \sigma_\beta^2, a, b) &\propto -\left (\frac{n}{2} + a  \right) +  \frac{e^{-\gamma}}{2} (\mathbf{y} - \mathbf{X}\boldsymbol\beta)^T(\mathbf{y} - \mathbf{X}\boldsymbol\beta) + b e^{-\gamma}. 
\end{aligned}
$$

We now have everything we need to solve the Hamiltonian equations via the leapfrog algorithm and generate samples for the posterior $f(\boldsymbol\theta)$. The main \texttt{hmc} function handles the details of the HMC sample generation process for the user. A description of the function parameters is in Section \textbf{A.1} of the Appendix. Additional programming details are provided with the \pkg{hmclearn} package, including detailed vignettes with additional examples.

For a numerical example we use the \textit{warpbreaks} dataset \citep{tippett1950technological}, which is one of the sample datasets included with base \proglang{R}. In this example, we estimate the associations between the yarn's type of wool and tension and the number of warp breaks per loom. We write the model as follows 
\begin{equation*}
\begin{split}
\text{Breaks}_i= &\beta_0+\beta_1 \text{woolB}_i + \beta_2 \text{tensionM}_i + \beta_3\text{tensionH}_i + \beta_4\text{woolB}_i:\text{tensionM}_i + \\
& \beta_5\text{woolB}_i:\text{tensionH}_i + \epsilon_i, 
\end{split}
\end{equation*}
where $y_i := \text{Breaks}_i$ and the \textit{i}th row of $\bold{X}$ is $\bold{x}_i^T = (1, \text{woolB}_i, \text{tensionM}_i, \text{tensionH}_i,\text{woolB}_i:\text{tensionM}_i, \text{woolB}_i:\text{tensionH}_i )$.

To fit this model using \texttt{hmc}, we must first specify the initial values of $\boldsymbol\theta$ for the MCMC chain. The initial values are provided as a vector of length $k = 7$, including 6 for $\boldsymbol\beta$ and 1 for $\gamma$. We use the default hyperparameters for the sample log posterior and gradient functions in \pkg{hmclearn}, such that $\sigma_\beta^2 = 1\mathrm{e}{3}$ and $a = b = 1\mathrm{e}{-4}$.   

The HMC simulation takes approximately 6 seconds to run on a 2015 Macbook Pro with a 2.5GHz processor. Users have a number of options to summarize and visualize the HMC samples. The generic \texttt{summary} function provides quantiles from the posterior samples in a table. Many data visualization options are available through direct integration with the \texttt{bayesplot} package  \citep{gabry2016bayesplot}. Graphical options for visualizing the posterior samples include histograms, density plots, and credible interval plots. General MCMC diagnostics such as trace plots, autocorrelation plots, and $\hat{R}$ statistics are also readily available. Additional customized analyses can be performed using the posterior sample output from \texttt{hmc}. 

The marginal posterior sample distributions for $f(\boldsymbol\theta)$ are found to be well-behaved and similar to frequentist estimates. The \proglang{R} code for fitting the model is presented in Section \textbf{A.2} of the Appendix.

\hypertarget{logistic-regression}{%
\subsubsection{Example 2: Logistic Regression}\label{logistic-regression}}
 
We consider a logistic regression model
$
\begin{aligned}
P(y_i = 1 | \bold{x}_i, \boldsymbol\beta) = \left[1 + \exp(-\bold{x}_i^T\boldsymbol\beta)\right]^{-1},
\end{aligned}
$
where $y_i$ is the binary response for the \textit{i}th subject $i=1,\dots, n$,  and $\bold{y}=(y_1, ..., y_n)^T $ is a vector of responses for all subjects.  The covariate values for the \textit{i}th subject are $\bold{x}_i^T = (x_{i0}, ..., x_{iq})$, where $x_{i0}$ is frequently set to one as an intercept term for all subjects. Frequently, $x_{i0}$ is set to one for all individuals as an intercept term.  We write the full design matrix as $\bold{X} = (\bold{x}_1^T, ..., \bold{x}_n^T)^T \in \mathbb{R}^{n\times(q+1)}$.  The regression coefficients for $q$ covariates plus an intercept are a vector $\boldsymbol\beta = (\beta_0, ..., \beta_q)^T$.

 The log likelihood for the logistic regression model is
$$
\begin{aligned}
\log f(\bold{y} | \bold{X}, \boldsymbol\beta) &=  \boldsymbol\beta^T\mathbf{X}^T(\mathbf{y} - \mathbf{1}_n) -\mathbf{1}_n^T [\log(1 + e^{-\mathbf{x}_i^T\boldsymbol\beta})]_{n\times 1} ,
\end{aligned}
$$
where $\boldsymbol\beta$ is the regression coefficient vector and the parameter we intend to estimate, and $[\log(1 + e^{-\mathbf{x}_i^T\boldsymbol\beta})]_{n\times 1}$ indicates an $n \times 1$ vector $\forall i = 1, \dots, n$. We specify a multivariate normal prior for $\boldsymbol\beta$ with covariance matrix $\sigma_\beta^2 \bold{I}$, where $\sigma_\beta^2$ is a hyperparameter set by the analyst.  

The log posterior is proportional to the sum of the log likelihood and log prior of $\boldsymbol\beta$. Excluding constants, we write the log posterior as
$$
\begin{aligned}
\log f(\boldsymbol\beta | \bold{y}, \bold{X}, \sigma_\beta^2) &\propto  \boldsymbol\beta^T\mathbf{X}^T(\mathbf{y} - \mathbf{1}_n) -\mathbf{1}_n^T [\log(1 + e^{-\mathbf{x}_i^T\boldsymbol\beta})]_{n\times 1} -\frac{\boldsymbol\beta^T \boldsymbol\beta}{2\sigma_\beta^2} .
\end{aligned}
$$

The parameters of interest are defined as $\boldsymbol\theta := \boldsymbol\beta = (\beta_0, ..., \beta_{q})^T$, where $k = q+1$. To fit this model using \texttt{hmc}, the user must provide a log posterior function containing the parameters of interest $\boldsymbol\theta$, the observed data, and possibly additional hyperparameters. The log posterior function for this model and the specification of priors are described in \pkg{hmclearn}. 

The Hamiltonian function (\ref{eq:Hamiltonian}) is composed of the log posterior and the log density function of the momentum $\bold{p} \sim N_k(0, \bold{M})$. Writing the Hamiltonian function for our example model is straightforward once the log posterior is specified,  
$$
\begin{aligned}
H(\boldsymbol\theta, \bold{p}) &= H(\boldsymbol\beta, \bold{p}) \propto  \log f(\boldsymbol\beta | \bold{y}, \bold{X}, \sigma_\beta^2) + \frac{1}{2}\bold{p}^T \bold{M}^{-1} \bold{p}.
\end{aligned}
$$

With the Hamiltonian function explicitly defined, we can write the Hamiltonian equations for this particular model. To generate samples from $f(\boldsymbol\theta)$, we then use the leapfrog method to find a discrete approximation.  The leapfrog steps  are integrated with \texttt{hmc} in a self-contained function, using user-supplied gradients. 
$$
\begin{aligned}
\nabla_{\boldsymbol\beta} \log f(\boldsymbol\beta  | \bold{y}, \bold{X}, \sigma_\beta^2) &\propto \mathbf{X}^T \left(\mathbf{y} - \mathbf{1}_n + \left [ \frac{e^{-\mathbf{x}_i^T \boldsymbol\beta}}{1 + e^{-\mathbf{x}_i^T\boldsymbol\beta}} \right ]_{n \times 1}  \right ) - \boldsymbol\beta / \sigma_\beta^2. 
\end{aligned}
$$

With the gradient function specified, we can solve the Hamiltonian equations via the leapfrog algorithm, and generate posterior samples following $f(\boldsymbol\theta)$. The main function \texttt{hmc}  handles the implementation of the HMC sample generation process. 

We analyzed data of 189 births at a U.S. hospital  \cite{hosmer1989multiple} to examine the risk factors of low birth weight. Data are available from the \texttt{MASS} package \citep{venables2013modern}. We prepare the data for analysis as noted in the text.

The logistic regression model formulation for this application is 
\begin{equation*}
\begin{split}
\text{logit}\left[P(\text{low}_i=1)\right]= &\beta_0+\beta_1 \text{age}_i + \beta_2 \text{lwt}_i + \beta_3\text{race2black}_i + \beta_4\text{race2other} +  \beta_5\text{smoke}_i +  \beta_6\text{ptd}_i +\\
& \beta_7\text{ht}_i + \beta_8\text{ui}_i + \beta_9\text{ftv21}_i + \beta_{10}\text{ftv22plus}_i.
\end{split}
\end{equation*}
Here $\bold{x}_i^T = (1,  \text{age}_i, \text{lwt}_i, \text{race2black}_i, \text{race2other}_i,  \text{smoke}_i, \text{ptd}_i, \text{ht}_i, \text{ui}_i, \text{ftv21}_i, \text{ftv22plus}_i)$, where
the elements indicate the mother's age in years \textit{age}, mother's weight in pounds at last menstrual period \textit{lwt}, black \textit{race2black} and other races \textit{race2other}, smoking during pregnancy \textit{smoke}, premature birth \textit{ptd}, hypertension \textit{ht}, presence of uterine irritability \textit{ui}, one physician visit during the first trimester \textit{ftv21}, and two or more physician visits during the first trimester \textit{ftv22plus}. 

To fit this model using \texttt{hmc}, the user needs to set the initial values for $\boldsymbol\beta$, a vector of length $k = 11$, as well as the value of the hyperparameter $\sigma_\beta^2 $, which we set at $1e3$. In this example, we set the step size parameter $\epsilon$ to different values for continuous and dichotomous variables.  

The HMC simulation takes about 6 seconds to run on a 2015 Macbook Pro with a 2.5GHz processor. The \proglang{R} code for fitting the model is presented in Section \textbf{A.3} of the Appendix. The marginal posterior sample distributions for $f(\boldsymbol\theta)$ are found to be well-behaved with central locations similar to frequentist estimates.

\hypertarget{mixed-effects-model}{%
\subsubsection{Example 3: Poisson regression with random subject effects}\label{mixed-effects-model}}

Finally, we consider a random effect model for count data 
$$
g[E(\bold{y}_i | u_i)] = \bold{X}_{i} \boldsymbol\beta + \bold{z}_{i} u_{i}, 
$$
for $i = 1, ..., n$ subjects, where each subject's response vector $\bold{y}_i = (y_{i1}, ..., y_{id})^T$ contains $j = 1, ..., d$ observations. Each individual has a subject-specific random intercept parameter $u_i$, and $\bold{u} = (u_1, ..., u_n)^T$. The fixed effects design matrix $\bold{X}_i = (\bold{x}_{i1}^T, ..., \bold{x}_{id}^T)^T \in \mathbb{R}^{d\times(q+1)}$, where the \textit{j}th row of $\bold{X}_i$ contains the $q+1$ covariate values of that observation, including a common intercept. The fixed effects regression coefficients for $q$ covariates and a global intercept are a vector $\boldsymbol\beta = (\beta_0, ..., \beta_q)^T$. The random intercept vector is $\bold{z}_i  = (z_{i1}, ..., z_{id})^T = \bold{1}_d$, The distribution of $\bold{y}_{i}$ conditional on $u_i$ follows a Poisson distribution with a log link function, where $\log [E(\bold{y}_i | u_i)] = \bold{X}_{i}\boldsymbol\beta + \bold{z}_{i} u_{j}$.

The subject-level response vectors are combined in a single vector, $\bold{y} = (\bold{y}_i^T, ..., \bold{y}_n^T)^T \in \mathbb{R}^{nd \times 1}$. The full fixed-effects design matrix for all subjects is $\bold{X} = (\bold{X}_1, ..., \bold{X}_n)^T \in \mathbb{R}^{nd\times (q+1)}$, and the random effects design matrix is $\bold{Z} = \bold{I}_n \otimes \bold{1}_d \in \mathbb{R}^{nd \times n}$. The log likelihood for the Poisson mixed effects model, omitting constants, can be written as
$$
\begin{aligned}
\log f(\bold{y} | \bold{X}, \bold{Z}, \boldsymbol\beta, \bold{u}) &\propto -\mathbf{1}_{nd}^T \left [ e^{\mathbf{x}_{ij}^T\boldsymbol\beta + z_{ij} u_i} \right ]_{nd \times 1}+ \mathbf{y}^T (\mathbf{X}\boldsymbol\beta + \mathbf{Z}\bold{u}), 
\end{aligned}
$$
where $\boldsymbol\beta$ is the fixed-effect coefficient vector, $u_i$ is the random intercept, and $ \left [ e^{\mathbf{x}_{ij}^T\boldsymbol\beta + z_{ij} u_i} \right ]_{nd \times 1}$ is an $nd \times 1$ vector $\forall i = 1, \dots, n \text{ and } j = 1, \dots d$. We specify multivariate normal priors $\boldsymbol\beta | \sigma_\beta^2 \sim N(0, \sigma_\beta^2 \bold{I})$ and $\bold{u} \sim N(0, \bold{G})$, where $\sigma_\beta^2$ is a hyperparameter set by the analyst and $\bold{G}$ is parameterized for efficient Bayesian computation. 

We parameterize the covariance matrix of $\bold{G}$ for efficient sampling of hierarchical models such that $\bold{G}^{1/2} := \lambda \bold{I} \boldsymbol\tau$, where $\boldsymbol\tau = (\tau_1, ..., \tau_n)^T \sim N(0, \bold{I}_n)$ \citep{betancourt_hamiltonian_2013}. For $\lambda$, we assign a 2-parameter half-t prior per the recommendation of \cite{gelman2006prior} for hierarchical models.

One final parameter transformation is necessary before applying HMC. Since the support of $\lambda$ is $(0, \infty)$, we apply a logarithmic transformation to expand the support to $\mathbb{R}$. We write
$$
\begin{aligned}
\xi &= \log \lambda, \quad
\lambda = g^{-1}(\xi) = e^\xi, \\
f(\xi | a, b) 
&\propto \left(1 + \frac{1}{\nu_\xi}\left(\frac{e^\xi}{A_\xi} \right)^2 \right)^{-(\nu_\xi + 1)/2} e^{\xi}, \\
\log f(\xi | a, b) &\propto-\frac{\nu_\xi + 1}{2} \log \left( 1 + \frac{1}{\nu_\xi} \left(\frac{e^\xi}{A_\xi} \right)^2 \right) + \xi, 
\end{aligned}
$$
where $\nu_\xi$ and $A_\xi$ are hyperparameters set by the analyst.

Omitting constants, we write the log posterior as 
$$
\begin{aligned}
\log f(\boldsymbol\beta,  \boldsymbol\tau, \xi | \bold{y}, \bold{X}, \bold{Z}, \sigma_\beta^2, \nu_\xi, A_\xi) &\propto-\mathbf{1}_{nd}^T \left [ e^{\mathbf{x}_{ij}^T\boldsymbol\beta + e^\xi z_{ij} \tau_i} \right ]_{nd \times 1} + \mathbf{y}^T (\mathbf{X}\boldsymbol\beta + e^\xi\mathbf{Z}\boldsymbol\tau) -\frac{\boldsymbol\beta^T \boldsymbol\beta}{2\sigma_\beta^2} - \\
&\qquad \frac{\nu_\xi + 1}{2} \log \left( 1 + \frac{1}{\nu_\xi} \left(\frac{e^\xi}{A_\xi} \right)^2 \right) + \xi-\frac{1}{2}\boldsymbol\tau^T \boldsymbol\tau,
\end{aligned}
$$
where the parameters of interest can be written as $\boldsymbol\theta := (\beta_0, ..., \beta_q, \tau_1, ..., \tau_n, \xi)^T$, with $k = q + n + 2$. 

Assuming $\bold{p} \sim N_k(0, \bold{M})$, we write the Hamiltonian function as, 
$$
\begin{aligned}
H(\boldsymbol\theta, \bold{p}) &= H(\boldsymbol\beta, \boldsymbol\tau, \xi, \bold{p}) \propto \log f(\boldsymbol\beta,  \boldsymbol\tau, \xi | \bold{y}, \bold{X}, \bold{Z}, \sigma_\beta^2, \nu_\xi, A_\xi)+ \frac{1}{2}\bold{p}^T \bold{M}^{-1} \bold{p},
\end{aligned}
$$
from which we can derive the Hamiltonian equations, and then use the leapfrog method to find approximate solutions.

We write the gradient functions for readers' convenience,
$$
\begin{aligned}
\nabla_{\boldsymbol\beta} \log f(\boldsymbol\beta, \xi, \boldsymbol\tau | \mathbf{y}, \mathbf{X}, \mathbf{Z}, \sigma_\beta^2, \nu_\xi, A_\xi)
&\propto \mathbf{X}^T \left ( - \left [ e^{\mathbf{x}_{ij}^T\boldsymbol\beta + e^\xi z_{ij} \tau_i} \right ]_{nd \times 1} + \mathbf{y}\right ) - \boldsymbol\beta/\sigma_\beta^2,    \\
\nabla_\xi \log f(\boldsymbol\beta, \xi, \boldsymbol\tau | \mathbf{y}, \mathbf{X}, \mathbf{Z}, \sigma_\beta^2, \nu_\xi, A_\xi)
&\propto e^\xi\boldsymbol\tau^T \mathbf{Z}^T \left (- \left [ e^{\mathbf{x}_{ij}^T\boldsymbol\beta + e^\xi z_{ij} \tau_i} \right ]_{nd \times 1}+  \mathbf{y} \right )- \frac{\nu_\xi + 1}{1 + \nu_\xi A_\xi^2 e^{-2\xi}} + 1, \\
\nabla_{\boldsymbol\tau} \log f(\boldsymbol\beta, \xi, \boldsymbol\tau | \mathbf{y}, \mathbf{X}, \mathbf{Z}, \sigma_\beta^2, \nu_\xi, A_\xi) &\propto  e^\xi \mathbf{Z}^T \left ( -\left [ e^{\mathbf{x}_{ij}^T\boldsymbol\beta + e^\xi z_{ij} \tau_i} \right ]_{nd \times 1}+  \mathbf{y} \right )- \boldsymbol\tau.
\end{aligned}
$$

For numerical example, we consider data generated by a study on gopher tortoises \citep{ozgul2009upper,fox2015ecological,bolkergithub}. The mortality of the tortoise populations is measured by the number of shells. We estimate the associations of the number of shells to year (2004, 2005, 2006) and seroprevalence of bacterium \emph{Mycoplasma agassizii}. The random effects are the intercepts for each of $n=10$ sites in Florida. Each site has $d=3$ observations, one for each year. The fixed effects are a global intercept, two indicator variables for the three years, and seroprevalence of \emph{M. agassizii}. 

The poisson mixed effects model can be written as 
\begin{equation}
\begin{split}
\log[E(\textbf{shells})] &\propto \sum_{i=1}^{10} \sum_{j=1}^3 \left [- e^{[1, I(2005)_{ij}, I(2006)_{ij}, \text{prev}_{ij}]\boldsymbol\beta + e^{\xi}z_{ij}\tau_i}+  \right. \\
&\qquad \left. y_{ij} \left( [1, I(2005)_{ij}, I(2006)_{ij}, \text{prev}_{ij}] \boldsymbol\beta + e^{\xi} z_{ij}\tau_i \right) \right ] - \\
&\quad  \frac{\boldsymbol\beta^T \boldsymbol\beta}{2\sigma_\beta^2} - 
\frac{\nu_\xi + 1}{2} \log \left( 1 + \frac{1}{\nu_\xi} \left(\frac{e^\xi}{A_\xi} \right)^2 \right) + \xi-\frac{1}{2}\boldsymbol\tau^T \boldsymbol\tau,
\end{split}
\end{equation}

where $\bold{y} := (\textbf{shells}_1, ...., \textbf{shells}_{10})^T$ and $\textbf{shells}_i = (\text{shells}_1, \text{shells}_2, \text{shells}_3)^T$. The fixed effects design matrix is composed from $\bold{x}_{ij}^T = [1, I(2005)_{ij}, I(2006)_{ij}, \text{prev}_{ij}]$, and the random effects design matrix from $z_{ij} = 1$ for $\text{site} \; i$ and 0 otherwise, for all observations $j = 1, 2, 3$.

To fit this model using \texttt{hmc}, we first specify the initial values of $\boldsymbol\theta$ in a vector of length $k= 15$ and use the default hyperparameters  $\sigma_\beta^2 = 1e3, \nu_{\xi}=1, \text{and } A_{\xi}=25$. The step sizes are selected as part of the tuning process.

The HMC simulation takes about 23 seconds to run on a 2015 Macbook Pro with a 2.5GHz processor. The marginal posterior sample distributions for $f(\boldsymbol\theta)$ are found to be well-behaved with central locations similar to frequentist estimates. The \proglang{R} code for fitting the model is presented in Section \textbf{A.4} of the Appendix.

In each of the above examples, we set $N = 2000$ HMC samples including a short burn-in period. The $\hat{R}$ statistics for each of the simulations is close to one, indicating that multiple chains converged to the same distribution for each example. Informally, the relatively low number of HMC simulations illustrates the efficiency benefits of this algorithm over traditional MCMC methods, such as the Metropolis algorithm, which often require many thousands of simulations to achieve a converge. A substantially larger number of simulations can push Metropolis to have a longer runtime than \textit{hmc} in \textit{hmclearn}, even when Metropolis is programmed in an efficient compiled language like \proglang{C++}.

\section{Discussion}

Since its becoming of a general-purpose computational method in the early 1990s, MCMC has fundamentally changed the landscape of Bayesian data analysis \citep{robert2011short}. Previous confinement to the conjugate families of distributions has been lifted, and analysts have been freed from the burden of explicitly deriving the posteriors. Over the past three decades, tremendous progress has been made in refining the MCMC methods, models are becoming more flexible, algorithms more comprehensive, and software easier to use. Despite the progress, however, as analysts begin to take on increasingly complex statistical models, suboptimal efficiency has become a predominant concern, especially in models involving high dimensional parameters. In many of those situations, the traditional MCMC is often too slow to be practically useful. 

One of the newer variants of MCMC algorithms designed to address the efficiency problem is HMC. With the aid of the posterior gradient functions and the Hamiltonian equations, HMC tends to converge to regions of higher posterior density more quickly in comparison with Metropolis-Hastings. For example, Section 5.7.1 of \cite{agresti_foundations_2015} uses \texttt{MCMCpack} \citep{mcmcpack2011} to fit a logistic regression model using Metropolis-Hastings.  The compiled \proglang{C++} code from this package is computationally advantageous compared to the fully \proglang{R}-based \pkg{hmclearn}. However, the run-time of this example with \pkg{MCMCpack} is approximately 2.6 minutes on a 2015 Macbook Pro with a 2.5GHz processor, versus 40 seconds with \pkg{hmclearn} on the same computer, a 5x difference. The code for this example is detailed in the Logistic Regression vignette provided for \pkg{hmclearn} on CRAN. Analysts who require efficient HMC computation without the need for manually computing gradients or tuning parameters may consider \pkg{Stan} \citep{carpenter_stan:_2017} for practical use. \proglang{Stan} translates BUGS-like \citep{spiegelhalter_bugs:_1999} code to \proglang{C++} code for efficient computation.

These exciting developments, however, have not been translated into analytical practice. Many statistical practitioners remain unfamiliar with these powerful tools and, thus, hesitant to use them. Some have attempted to generate HMC samples by mimicking the \proglang{Stan} code, but in the absence of an in-depth understanding of the method and the ideas behind it, many analysts have not acquired a level of comfort to write HMC code for less standard analyses. We contend that the best way to learn a new method is through hands-on data analysis, with common statistical models on a familiar computational platform. With this in mind, we have put forward an introductory level description of HMC, not with the original terminology of classical mechanics, but in a more familiar language of statistics. We have disseminated the components of the HMC algorithm and discussed the implementation details, from prior specification, posterior and gradient function derivation, to solving the Hamiltonian differential equations, and to the tuning of HMC parameters. Herein, we present an \proglang{R} package \pkg{hmclearn} to help beginners to experiment with HMC in a familiar computing environment. The main function of this package, \texttt{hmc} is designed for general use -- analysts could use it to produce MCMC samples by using user-supplied posterior functions. We have provided many concrete data examples, in the package as well as in this manuscript, to help learners study and appreciate the inner workings of the algorithm. In comparison with commonly used Bayesian data analysis software such as \proglang{Stan}, our package \pkg{hmclearn} is designed primarily as a teaching tool. As such, the input functions require hands-on programming, so that the data generation process is made more transparent to its users. This said, we would not trivialize the potential challenges in implementing a successful HMC program. The tuning of parameters, for example, often requires much practice and experience. Notwithstanding such limitations, we hope that this paper provides an intuitive introduction of a powerful and yet intricate computational tool.

\pagebreak

\section{Supplementary Materials}

\begin{description}

\item[Appendix:] R code for HMC examples. (pdf file type)

\item[R-package for learning HMC:] R-package \pkg{hmclearn} contains a general-purpose function as well as utility functions for the model fitting methods described in the article. Example data sets and code are also made available in the package. The package \pkg{hmclearn} can be accessed at \url{https://cran.r-project.org/web/packages/hmclearn/index.html}.

\pagebreak

\end{description}

\bibliographystyle{agsm}

\bibliography{hmc}
\end{document}


\articletype{Supplementary Material}

\title{Appendices to Learning Hamiltonian Monte Carlo in R}

\author{\name{Samuel Thomas$^{a}$, Wanzhu Tu$^{a}$}
\affil{$^{a}$Indiana University School of Medicine, 410 W 10th St \#3000,
Indianapolis, IN 46202}
}


\maketitle

\begin{abstract}
The supplementary material contains a short description of the the R package \textbf{hmclearn}, and code for three examples described in the main paper.
\end{abstract}


\appendix
\section{\textit{hmclearn} package}

\subsection{Main function:  \textit{hmc} }

R package \textbf{hmclearn} contains the functions for generating Hamiltonian Monte Carlo (HMC) samples. To download \textbf{hmclearn}, go to \url{https://cran.r-project.org/web/packages/hmclearn/index.html}.

The main function of \textbf{hmclearn}  is \textit{hmc}, which is a general-purpose function for producing posterior samples with user-supplied log-posterior and gradient functions. Most of the input parameters of \textit{hmc} have preset values. These defaults which can be overwritten by  custom values.

\begin{itemize}
\item \textit{N}:  Number of MCMC samples.  Default is 10,000.
\item \textit{theta.init}:  Initial values for model parameters.
\item \textit{epsilon}:  Stepsize tuning parameter for HMC. Default is 0.01.
\item \textit{L}:  Number of Leapfrog steps tuning parameter for HMC. Default is 10. 
\item \textit{logPOSTERIOR}:  Function to return the log posterior depending on $\boldsymbol\theta$, hyperparameters, and data all provided in a list passed to \textit{param}.
\item \textit{glogPOSTERIOR}:  Function to return the gradient of the log posterior depending on $\boldsymbol\theta$, hyperparameters, and data all provided in a list passed to \textit{param}.
\item \textit{varnames}: Optional vector of variable names in the model. This is used for \textit{summary} and plotting functions.
\item \textit{randlength}:  Logical indicator on whether to apply some randomness to the number of Leapfrog steps \textit{L}. Default is FALSE.  
\item \textit{Mdiag}:  Optional vector for the diagonal of the Mass matrix $M$. The default is the identity matrix.
\item \textit{constrain}:  Optional vector of which variables are bounded as positive only. The leapfrog routine is adjusted from the default for these parameters. Default is FALSE for all parameters.
\item \textit{verbose}:  Logical indicator on whether to print status updates of \textit{hmc}. Default is FALSE.
\item \textit{param}:  List of data objects and hyperparameters passed to \textit{logPOSTERIOR} and \textit{glogPOSTERIOR}.
\item \textit{chains}: The number of HMC chains to run. The default is 1, although multiple chains are recommended. 
\item \textit{parallel}: Logical indicator on whether to run parallel processing for multiple chains. Setting this parameter to TRUE can significantly speedup computation, but there can be technical complications depending on the user's platform. Default is FALSE, which means that multiple chains are run sequentially.
\end{itemize}

Function \textit{hmc} requires users to provide log posterior function and its gradient,  \textit{logPOSTERIOR} and \textit{glogPOSTERIOR}. The prior can be specified as part of the log posterior.  Data can be directly input into these functions, or the objects (e.g. $y$, $X$, and $Z$) can be passed in a list provided to \textit{param}. The purpose of this design is to maximize flexibility regarding the types of models.

The default number of simulations is set to $10,000$. However, the user may opt to start with a smaller number during tuning. Well-tuned models tend to require fewer MCMC samples. Initial values for the parameters must be provided by the user. Caution should be exercised to ensure that the initial values are in the support of \(\boldsymbol\theta\). 

For the interested reader, the main HMC algorithm is in the \textit{hmc.fit} function, while \textit{hmc} is a higher level function whose purpose is to govern serial or parallel processing and create the \textit{hmclearn} object.

Function \textit{hmc} outputs a single list containing the simulated samples accepted in the accept/reject step.

\subsection{Example 1: Linear Regression}

The dataset \textit{warpbreaks} is available standard with \textit{R}.

\begin{CodeChunk}
\begin{CodeInput}
R> head(warpbreaks)
\end{CodeInput}
\end{CodeChunk}

\begin{tabular}{rll}
\toprule
breaks & wool & tension\\
\midrule
26 & A & L\\
30 & A & L\\
54 & A & L\\
25 & A & L\\
70 & A & L\\
52 & A & L\\
\bottomrule
\end{tabular}

\begin{CodeChunk}
\begin{CodeInput}
R> summary(warpbreaks)
\end{CodeInput}
\end{CodeChunk}

\begin{tabular}{llll}
\toprule
  &     breaks & wool & tension\\
\midrule
 & Min.   :10.00 & A:27 & L:18\\
 & 1st Qu.:18.25 & B:27 & M:18\\
 & Median :26.00 &  & H:18\\
 & Mean   :28.15 &  & \\
 & 3rd Qu.:34.00 &  & \\
 & Max.   :70.00 &  & \\
\bottomrule
\end{tabular}
\\
\\
Variables of interest in this dataset are:

\begin{itemize}
\item breaks: the number of breaks (continuous)
\item woolB:  indicator for wool type B (0/1)
\item tensionM:  indicator for level of tension M (0/1)
\item tensionH:  indicator for level of tension H (0/1)
\item woolB:tensionM:  interaction of wool type B and tension level M (0/1)
\item woolB:tensionH:  interaction of wool type B and tension level H (0/1)
\end{itemize}

The dependent variable \textit{breaks} is stored in \(y\). The design
matrix can be constructed using standard \textit{model.matrix} function
in \textit{R}.

\begin{CodeChunk}
\begin{CodeInput}
R> y <- warpbreaks$breaks
R> X <- model.matrix(breaks ~ wool*tension, data=warpbreaks)
\end{CodeInput}
\end{CodeChunk}

The log posterior and gradient functions are based on the likelihood and
prior choices in this example.

\begin{Shaded}
\begin{Highlighting}[]
\NormalTok{linear_posterior <-}\StringTok{ }\ControlFlowTok{function}\NormalTok{(theta, y, X, }\DataTypeTok{a=}\FloatTok{1e-4}\NormalTok{, }\DataTypeTok{b=}\FloatTok{1e-4}\NormalTok{, }
                             \DataTypeTok{sig2beta=}\FloatTok{1e3}\NormalTok{) \{}
\NormalTok{  k <-}\StringTok{ }\KeywordTok{length}\NormalTok{(theta)}
\NormalTok{  beta_param <-}\StringTok{ }\KeywordTok{as.numeric}\NormalTok{(theta[}\DecValTok{1}\OperatorTok{:}\NormalTok{(k}\DecValTok{-1}\NormalTok{)])}
\NormalTok{  gamma_param <-}\StringTok{ }\NormalTok{theta[k]}

\NormalTok{  n <-}\StringTok{ }\KeywordTok{nrow}\NormalTok{(X)}
\NormalTok{  result <-}\StringTok{ }\OperatorTok{-}\NormalTok{(n}\OperatorTok{/}\DecValTok{2}\OperatorTok{+}\NormalTok{a)}\OperatorTok{*}\NormalTok{gamma_param }\OperatorTok{-}\StringTok{ }\KeywordTok{exp}\NormalTok{(}\OperatorTok{-}\NormalTok{gamma_param)}\OperatorTok{/}\DecValTok{2} \OperatorTok{*}\StringTok{ }
\StringTok{    }\KeywordTok{t}\NormalTok{(y }\OperatorTok{-}\StringTok{ }\NormalTok{X}\OperatorTok{%*%}\NormalTok{beta_param) }\OperatorTok{%*%}
\StringTok{    }\NormalTok{(y }\OperatorTok{-}\StringTok{ }\NormalTok{X}\OperatorTok{%*%}\NormalTok{beta_param) }\OperatorTok{-}\StringTok{ }\NormalTok{b}\OperatorTok{*}\KeywordTok{exp}\NormalTok{(}\OperatorTok{-}\NormalTok{gamma_param) }\OperatorTok{-}\StringTok{ }
\StringTok{    }\DecValTok{1}\OperatorTok{/}\DecValTok{2}\OperatorTok{*}\StringTok{ }\KeywordTok{t}\NormalTok{(beta_param) }\OperatorTok{%*%}\StringTok{ }\NormalTok{beta_param }\OperatorTok{/}\StringTok{ }\NormalTok{sig2beta}
  \KeywordTok{return}\NormalTok{(result)}
\NormalTok{\}}
\end{Highlighting}
\end{Shaded}

\begin{Shaded}
\begin{Highlighting}[]
\NormalTok{g_linear_posterior <-}\StringTok{ }\ControlFlowTok{function}\NormalTok{(theta, y, X, }\DataTypeTok{a=}\FloatTok{1e-4}\NormalTok{, }\DataTypeTok{b=}\FloatTok{1e-4}\NormalTok{, }
                             \DataTypeTok{sig2beta=}\FloatTok{1e3}\NormalTok{) \{}
\NormalTok{  k <-}\StringTok{ }\KeywordTok{length}\NormalTok{(theta)}
\NormalTok{  beta_param <-}\StringTok{ }\KeywordTok{as.numeric}\NormalTok{(theta[}\DecValTok{1}\OperatorTok{:}\NormalTok{(k}\DecValTok{-1}\NormalTok{)])}
\NormalTok{  gamma_param <-}\StringTok{ }\NormalTok{theta[k]}
\NormalTok{  n <-}\StringTok{ }\KeywordTok{nrow}\NormalTok{(X)}

\NormalTok{  grad_beta <-}\StringTok{ }\KeywordTok{exp}\NormalTok{(}\OperatorTok{-}\NormalTok{gamma_param)  }\OperatorTok{*}\StringTok{ }\KeywordTok{t}\NormalTok{(X) }\OperatorTok{%*%}\StringTok{ }
\StringTok{    }\NormalTok{(y }\OperatorTok{-}\StringTok{ }\NormalTok{X}\OperatorTok{%*%}\NormalTok{beta_param)  }\OperatorTok{-}\StringTok{ }\NormalTok{beta_param }\OperatorTok{/}\StringTok{ }\NormalTok{sig2beta}
\NormalTok{  grad_gamma <-}\StringTok{ }\OperatorTok{-}\NormalTok{(n}\OperatorTok{/}\DecValTok{2} \OperatorTok{+}\StringTok{ }\NormalTok{a) }\OperatorTok{+}\StringTok{ }\KeywordTok{exp}\NormalTok{(}\OperatorTok{-}\NormalTok{gamma_param)}\OperatorTok{/}\DecValTok{2} \OperatorTok{*}\StringTok{ }
\StringTok{    }\KeywordTok{t}\NormalTok{(y }\OperatorTok{-}\StringTok{ }\NormalTok{X}\OperatorTok{%*%}\NormalTok{beta_param) }\OperatorTok{%*%}
\StringTok{    }\NormalTok{(y }\OperatorTok{-}\StringTok{ }\NormalTok{X}\OperatorTok{%*%}\NormalTok{beta_param) }\OperatorTok{+}\StringTok{ }\NormalTok{b}\OperatorTok{*}\KeywordTok{exp}\NormalTok{(}\OperatorTok{-}\NormalTok{gamma_param)}
  \KeywordTok{c}\NormalTok{(}\KeywordTok{as.numeric}\NormalTok{(grad_beta), }\KeywordTok{as.numeric}\NormalTok{(grad_gamma))}
\NormalTok{\}}
\end{Highlighting}
\end{Shaded}
The vector of parameters of interest is
\((\boldsymbol\beta, \gamma) \in \theta\). The initial values specified
in a vector of length 6 for \(\beta\) plus 1 for \(\gamma\). The step
size is a factor of 10 higher for \(\boldsymbol\beta\) than for log
transformed variance.

\begin{CodeChunk}
\begin{CodeInput}
R> N <- 2e3
R> set.seed(143)
R> 
R> eps_vals <- c(rep(2e-1, 6), 2e-2) 
R>
R> fm1_hmc <- hmc(N, theta.init = c(rep(0, 6), 1), 
R>				epsilon = eps_vals, L = 20, 
R>				logPOSTERIOR = linear_posterior, 
R>				glogPOSTERIOR = g_linear_posterior, 
R>				varnames = c(colnames(X), "log_sigma_sq"), 
R>				param = list(y = y, X = X), chains = 2, 
R>				parallel = FALSE)
\end{CodeInput}
\end{CodeChunk}

The average acceptance rate over the two chains is $0.96$, which is appropriate for a relatively simple model such as this one. The \textit{parallel} parameter is set to FALSE to run each MCMC chain sequentially. This parameter can be set to TRUE to run both MCMC chains at the same time on multiple cores.

We summarize the results and plot the histograms of the simulated posteriors. The $\hat{R}$ statistics are all close to one, indicating that both chains converged to the same distribution. 

\begin{CodeChunk}
\begin{CodeInput}
R> summary(fm1_hmc, burnin=200)
\end{CodeInput}
\end{CodeChunk}

\begin{figure}
\begin{center}
\includegraphics[width=0.9\textwidth]{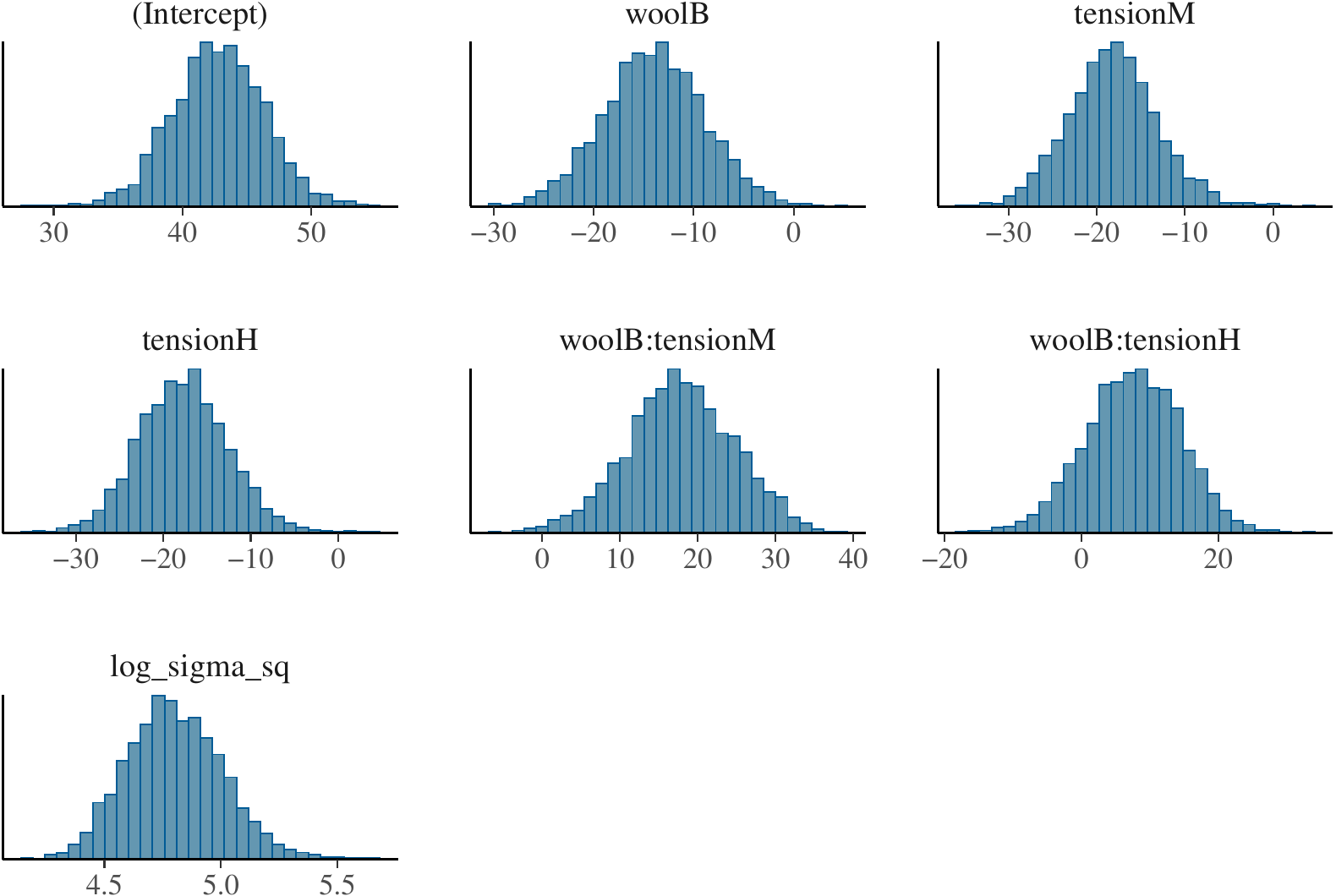}
\end{center}
\caption{Histograms from marginal posteriors for example 1.}
\label{fig:hist1}
\end{figure}

\resizebox{1 \textwidth}{!}{
\begin{tabular}{lrrrrrrrr}
\toprule
  & 2.5\% & 5\% & 25\% & 50\% & 75\% & 95\% & 97.5\% & rhat\\
\midrule
(Intercept) & 35.733 & 37.010 & 40.464 & 42.801 & 45.197 & 48.346 & 49.533 & 1.004\\
woolB & -23.693 & -22.183 & -17.223 & -13.945 & -10.517 & -5.763 & -3.986 & 1.002\\
tensionM & -27.737 & -26.290 & -21.391 & -18.194 & -14.962 & -9.883 & -8.052 & 1.006\\
tensionH & -27.473 & -25.884 & -21.214 & -17.708 & -14.388 & -9.505 & -7.604 & 1.001\\
woolB:tensionM & 4.086 & 6.469 & 13.301 & 17.717 & 22.481 & 29.030 & 31.019 & 1.006\\
woolB:tensionH & -6.217 & -4.112 & 3.029 & 7.709 & 12.492 & 18.525 & 20.394 & 1.000\\
log\_sigma\_sq & 4.425 & 4.484 & 4.659 & 4.793 & 4.937 & 5.144 & 5.218 & 1.000\\
\bottomrule
\end{tabular}
}
\\
\\
The posterior distributions can be visualized by using the \textit{plot} function. See  Fig \ref{fig:hist1}.

\begin{CodeChunk}
\begin{CodeInput}
R> plot(fm1_hmc, burnin=200)
\end{CodeInput}
\end{CodeChunk}

Note that the Inverse Gamma distribution is not always an optimal prior when the support is strictly positive. This prior can lead to problematic results when the true value of the parameter is close to zero. Half-t distributions usually provide a more stable alternative to Inverse Gamma \citep{gelman2006prior}.

\begin{figure}
\begin{center}
\includegraphics[width=0.9\textwidth]{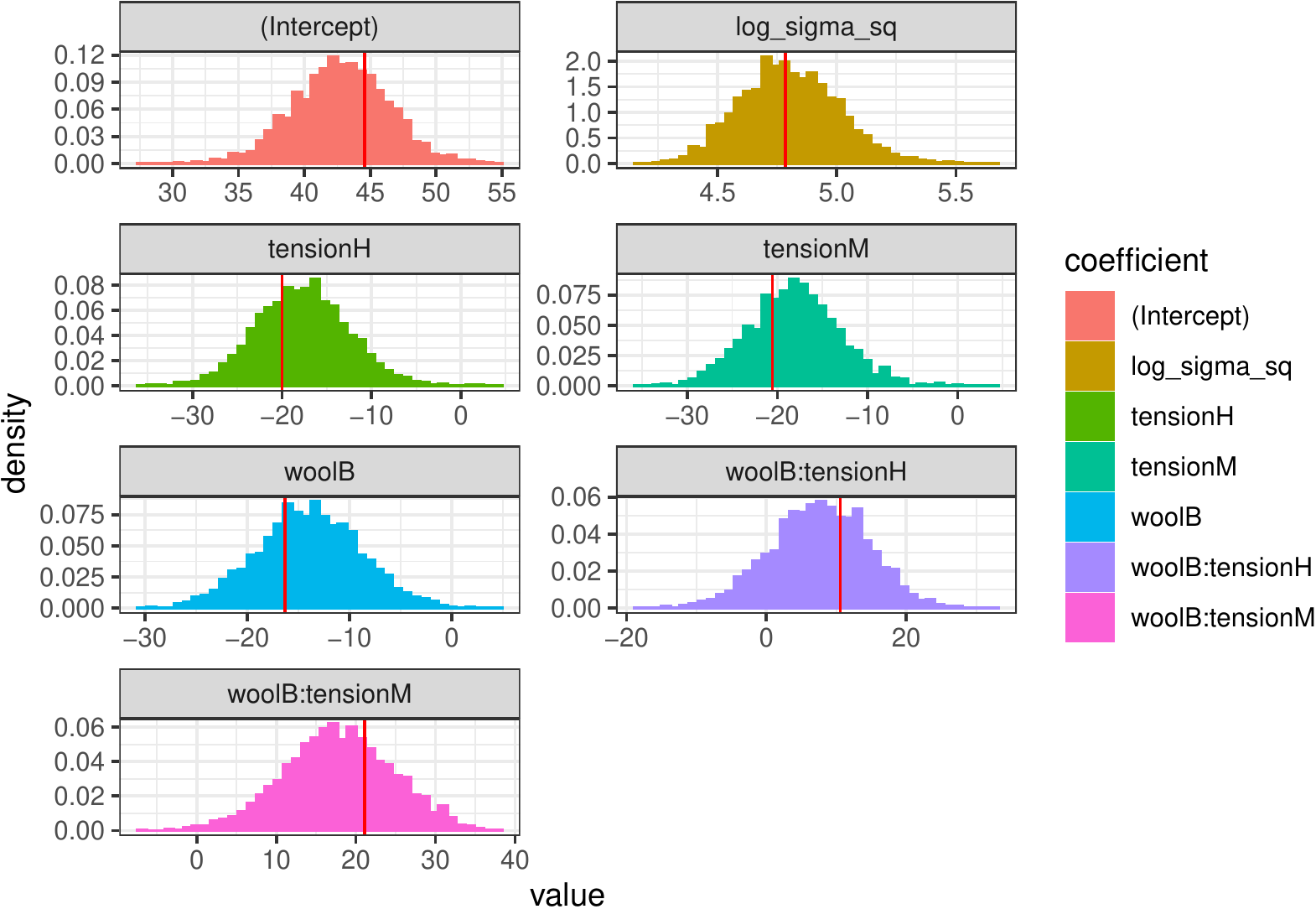}
\end{center}
\caption{Diagnostic plots comparing HMC (histograms) to frequentist (vertical lines) estimates for example 1. The HMC estimates are in line with those from \textit{lm}. }
\label{fig:diag1}
\end{figure}

In this example, the posterior estimates are comparable to frequentist estimates. See Fig \ref{fig:diag1}.

\begin{CodeChunk}
\begin{CodeInput}
R> f <- lm(breaks ~ wool*tension, data = warpbreaks)
R> freq.param <- c(coef(f), 2*log(sigma(f)))
R> diagplots(fm1_hmc, burnin=200, comparison.theta=freq.param)
\end{CodeInput}
\end{CodeChunk}

\subsection{Example 2: Logistic Regression}

The data for this example is from a study of 189 births at a U.S. hospital  \citep{hosmer1989multiple}. The dependent variable is an indicator of low birth weight. Data is available from the MASS package \citep{venables2013modern}. We prepare the data for analysis as noted in the main text.

\begin{CodeChunk}
\begin{CodeInput}
R> birthwt2 <- MASS::birthwt
R> birthwt2$race2 <- factor(birthwt2$race, 
			labels = c("white", "black", "other"))
R> birthwt2$ptd <- ifelse(birthwt2$ptl > 0, 1, 0)
R> birthwt2$ftv2 <- factor(ifelse(birthwt2$ftv > 2, 2, birthwt2$ftv),
			labels = c("0", "1", "2+"))
R> X <- model.matrix(low ~ age + lwt + race2 + smoke + 
			ptd + ht + ui + ftv2, 
			data = birthwt2)
R> y <- birthwt2$low
\end{CodeInput}
\end{CodeChunk}

Variables of interest in this dataset are:

\begin{itemize}
\item low:  birth weight less than 2.5kg (0/1)
\item age:  age of mother (yrs)
\item lwt:  weight of mother (lbs)
\item race2:  factor white/black/other
\item smoke:  smoking indicator (0/1)
\item ptd:  premature labor indicator (0/1)
\item ht:  history of hypertension indicator (0/1)
\item ui:  uterine irritability indicator (0/1)
\item ftv2:  number of physician visits factor (0, 1, 2 or more)
\end{itemize}

Two of the independent variables are continuous with wide ranges of values. The other nine variables are all dichotomous. In tuning this model, the step size \(\epsilon\) is tuned separately to each of these types of variables. This example illustrates the need to set the tuning parameters for the specific HMC application.

The log posterior and gradient functions are based on the likelihood and prior choices in this example.

\begin{Shaded}
\begin{Highlighting}[]
\NormalTok{logistic_posterior <-}\StringTok{ }\ControlFlowTok{function}\NormalTok{(theta, y, X, }\DataTypeTok{sig2beta=}\FloatTok{1e3}\NormalTok{) \{}
\NormalTok{  k <-}\StringTok{ }\KeywordTok{length}\NormalTok{(theta)}
\NormalTok{  beta_param <-}\StringTok{ }\KeywordTok{as.numeric}\NormalTok{(theta)}
\NormalTok{  onev <-}\StringTok{ }\KeywordTok{rep}\NormalTok{(}\DecValTok{1}\NormalTok{, }\KeywordTok{length}\NormalTok{(y))}

\NormalTok{  ll_bin <-}\StringTok{ }\KeywordTok{t}\NormalTok{(beta_param) }\OperatorTok{%*%}\StringTok{ }\KeywordTok{t}\NormalTok{(X) }\OperatorTok{%*%}\StringTok{ }\NormalTok{(y }\OperatorTok{-}\StringTok{ }\DecValTok{1}\NormalTok{) }\OperatorTok{-}
\StringTok{    }\KeywordTok{t}\NormalTok{(onev) }\OperatorTok{%*%}\StringTok{ }\KeywordTok{log}\NormalTok{(}\DecValTok{1} \OperatorTok{+}\StringTok{ }\KeywordTok{exp}\NormalTok{(}\OperatorTok{-}\NormalTok{X }\OperatorTok{%*%}\StringTok{ }\NormalTok{beta_param))}

\NormalTok{  result <-}\StringTok{ }\NormalTok{ll_bin }\OperatorTok{-}\StringTok{ }\DecValTok{1}\OperatorTok{/}\DecValTok{2}\OperatorTok{*}\StringTok{ }\KeywordTok{t}\NormalTok{(beta_param) }\OperatorTok{%*%}\StringTok{ }
\StringTok{    }\NormalTok{beta_param }\OperatorTok{/}\StringTok{ }\NormalTok{sig2beta}

  \KeywordTok{return}\NormalTok{(result)}
\NormalTok{\}}
\end{Highlighting}
\end{Shaded}

\begin{Shaded}
\begin{Highlighting}[]
\NormalTok{g_logistic_posterior <-}\StringTok{ }\ControlFlowTok{function}\NormalTok{(theta, y, X, }\DataTypeTok{sig2beta=}\FloatTok{1e3}\NormalTok{) \{}
\NormalTok{  n <-}\StringTok{ }\KeywordTok{length}\NormalTok{(y)}
\NormalTok{  k <-}\StringTok{ }\KeywordTok{length}\NormalTok{(theta)}
\NormalTok{  beta_param <-}\StringTok{ }\KeywordTok{as.numeric}\NormalTok{(theta)}

\NormalTok{  result <-}\StringTok{ }\KeywordTok{t}\NormalTok{(X) }\OperatorTok{%*%}\StringTok{ }\NormalTok{( y }\OperatorTok{-}\StringTok{ }\DecValTok{1}  \OperatorTok{+}\StringTok{ }\KeywordTok{exp}\NormalTok{(}\OperatorTok{-}\NormalTok{X }\OperatorTok{%*%}\StringTok{ }\NormalTok{beta_param) }\OperatorTok{/}
\StringTok{              }\NormalTok{(}\DecValTok{1} \OperatorTok{+}\StringTok{ }\KeywordTok{exp}\NormalTok{(}\OperatorTok{-}\NormalTok{X }\OperatorTok{%*%}\StringTok{ }\NormalTok{beta_param))) }\OperatorTok{-}\NormalTok{beta_param}\OperatorTok{/}\NormalTok{sig2beta}

  \KeywordTok{return}\NormalTok{(result)}
\NormalTok{\}}
\end{Highlighting}
\end{Shaded}

We set the initial values to zero and hyperparameters to their default values.

\begin{CodeChunk}
\begin{CodeInput}
R> N <- 2e3
R> continuous_ind <- c(FALSE, TRUE, TRUE, rep(FALSE, 8))
R> eps_vals <- ifelse(continuous_ind, 1e-3, 5e-2)
R> 
R> set.seed(143)
R> fm2_hmc <- hmc(N, theta.init = rep(0, 11), 
				epsilon = eps_vals, L = 10, 
				logPOSTERIOR = logistic_posterior, 
				glogPOSTERIOR = g_logistic_posterior, 
				param = list(y = y, X = X), 
				varnames = colnames(X), 
				chains = 2, parallel = FALSE)
R> 
R> fm2_accept/N
\end{CodeInput}

\begin{CodeOutput}
 [1]  0.8105 0.8215
\end{CodeOutput}

\end{CodeChunk}

The average acceptance rate over two chains is $0.82$, which is appropriate for a relatively simple model such as this one. The \textit{parallel} parameter is set here to run each MCMC chain sequentially. This parameter can be set to \textit{TRUE} to run both MCMC chains at the same time on multiple cores.

We summarize the results and plot the histograms of the simulated posteriors. The $\hat{R}$ statistics are all close to one, indicating that both chains converged to the same distribution. 

\begin{CodeChunk}
\begin{CodeInput}
R> summary(fm2_hmc, burnin=200)
\end{CodeInput}
\end{CodeChunk}

\resizebox{1 \textwidth}{!}{
\begin{tabular}{lrrrrrrrr}
\toprule
  & 2.5\% & 5\% & 25\% & 50\% & 75\% & 95\% & 97.5\% & rhat\\
\midrule
(Intercept) & -1.464 & -1.093 & 0.240 & 1.150 & 2.000 & 3.134 & 3.557 & 1.008\\
age & -0.115 & -0.106 & -0.068 & -0.045 & -0.020 & 0.021 & 0.032 & 1.010\\
lwt & -0.032 & -0.030 & -0.022 & -0.017 & -0.012 & -0.006 & -0.004 & 1.005\\
race2black & 0.132 & 0.309 & 0.840 & 1.210 & 1.596 & 2.154 & 2.328 & 1.000\\
race2other & -0.167 & -0.029 & 0.423 & 0.737 & 1.063 & 1.561 & 1.708 & 1.001\\
smoke & -0.121 & 0.015 & 0.444 & 0.752 & 1.046 & 1.507 & 1.627 & 1.000\\
ptd & 0.530 & 0.676 & 1.151 & 1.474 & 1.813 & 2.293 & 2.452 & 1.000\\
ht & 0.544 & 0.807 & 1.573 & 2.061 & 2.553 & 3.300 & 3.535 & 1.005\\
ui & -0.267 & -0.117 & 0.373 & 0.685 & 1.018 & 1.469 & 1.594 & 1.000\\
ftv21 & -1.464 & -1.332 & -0.798 & -0.475 & -0.167 & 0.288 & 0.453 & 1.007\\
ftv22+ & -0.751 & -0.599 & -0.146 & 0.156 & 0.508 & 0.980 & 1.119 & 1.003\\
\bottomrule
\end{tabular}
}

\begin{figure}
\begin{center}
\includegraphics[width=0.9\textwidth]{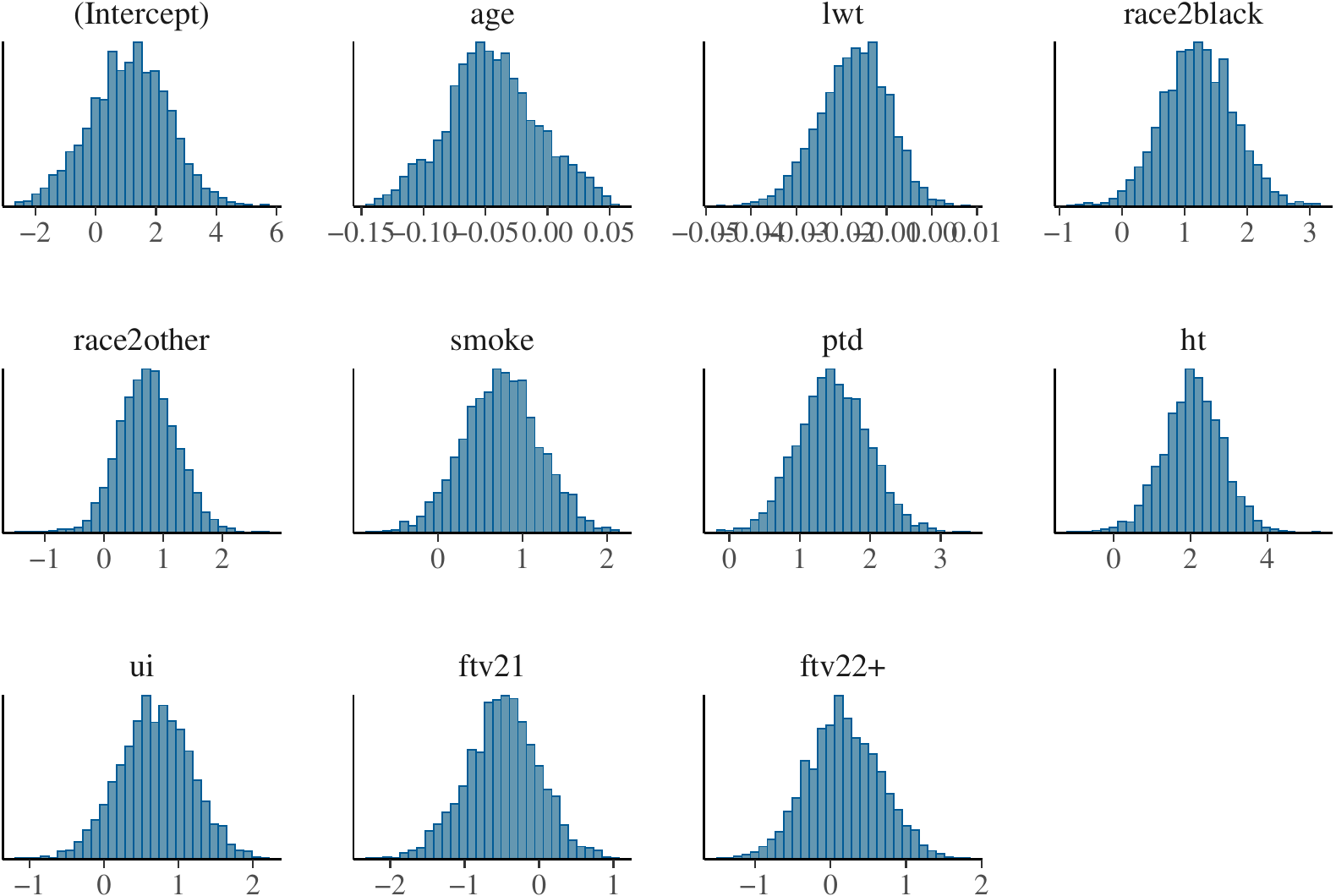}
\end{center}
\caption{Histograms from marginal posteriors for example 2.}
\label{fig:hist2}
\end{figure}

The distributions of the marginal posteriors can be visualized by using the \textit{plot} function. See  Fig \ref{fig:hist2}.

\begin{CodeChunk}
\begin{CodeInput}
R> plot(fm2_hmc, burnin=200)
\end{CodeInput}
\end{CodeChunk}

\begin{figure}
\begin{center}
\includegraphics[width=0.9\textwidth]{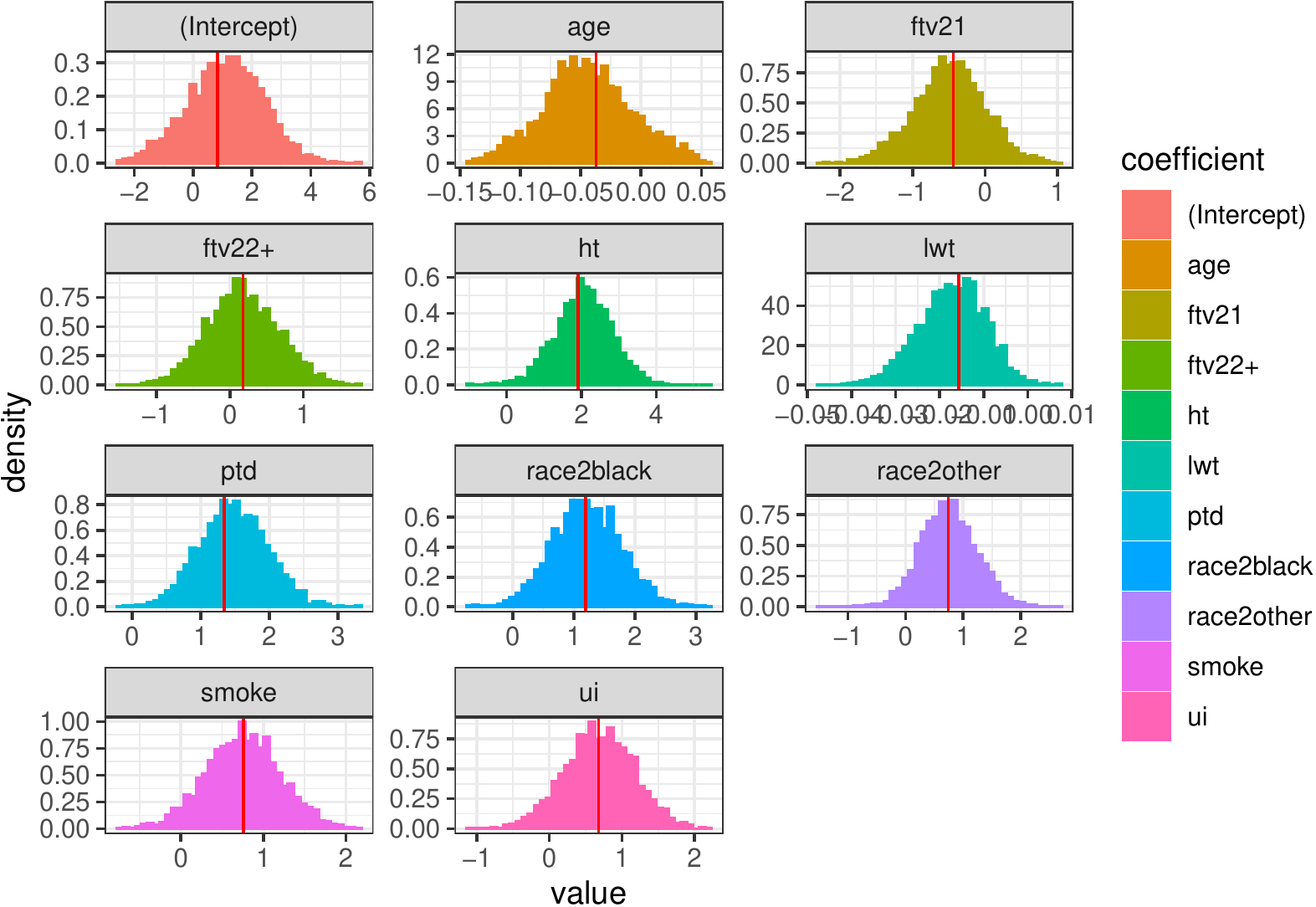}
\end{center}
\caption{Diagnostic plots comparing HMC (histograms) to frequentist (vertical lines) estimates for example 2. The HMC estimates are in line with those from \textit{glm}. }
\label{fig:diag2}
\end{figure}

In this example, the posterior estimates are comparable to frequentist estimates. See Fig \ref{fig:diag2}.

\begin{CodeChunk}
\begin{CodeInput}
R> f2 <- glm(low ~ age + lwt + race2 + smoke + ptd + ht + ui + ftv2, 
		data = birthwt2, family = binomial)
R> freq.param2 <- coef(f2)
R> diagplots(fm2_hmc, burnin=200, comparison.theta = freq.param2)
\end{CodeInput}
\end{CodeChunk}

\subsection{Example 3: Mixed effects Poisson regression model}

The data for this example is from a study on gopher tortoises \citep{ozgul2009upper,fox2015ecological,bolkergithub}.

Variables of interest are:
\begin{itemize}
\item shells:  count of shells
\item factor.year.2005:  indicator for year 2005 (0/1)
\item factor.year.2006:  indicator for year 2006 (0/1)
\item prev:  Seroprevalence to \textit{Mycoplasma agassizii} (continuous)
\end{itemize}

The design matrices \(X\) and \(Z\) must be set up for \textit{hmc}. The fixed effects matrix \(X\) contains a global intercept and covariates for year 2005 \textit{factor.year.2005}, year 2006 \textit{factor.year.2006}, and Seroprevalence \textit{prev}.

A random intercept is generated for each of the 10 sites in the dataset and stored as a block diagonal matrix in \(Z\).

\begin{CodeChunk}
\begin{CodeInput}
R> data(Gdat)
R> 
R> Zi.lst <- split(rep(1, nrow(Gdat)), Gdat$Site)
R> Zi.lst <- lapply(Zi.lst, as.matrix)
R> Z <- Matrix::bdiag(Zi.lst)
R> Z <- as.matrix(Z)
R> X <- model.matrix(~ factor(year), data = Gdat)
R> X <- cbind(X, Gdat$prev)
R> colnames(X)[ncol(X)] <- "prev"
R> colnames(X) <- make.names(colnames(X))
R> colnames(X)[1] <- "intercept"
R> y <- Gdat$shells
\end{CodeInput}
\end{CodeChunk}

The log posterior and gradient functions are based on the likelihood and prior choices in this example.

\begin{Shaded}
\begin{Highlighting}[]
\NormalTok{glmm_poisson_posterior <-}\StringTok{ }\ControlFlowTok{function}\NormalTok{(theta, y, X, Z, n, }\DataTypeTok{nrandom=}\DecValTok{1}\NormalTok{,}
                                   \DataTypeTok{nuxi=}\DecValTok{1}\NormalTok{, }\DataTypeTok{Axi=}\DecValTok{25}\NormalTok{, }\DataTypeTok{sig2beta=}\FloatTok{1e3}\NormalTok{) \{}
\NormalTok{  Z <-}\StringTok{ }\KeywordTok{as.matrix}\NormalTok{(Z)}
\NormalTok{  p <-}\StringTok{ }\KeywordTok{ncol}\NormalTok{(X)}

\NormalTok{  beta_param <-}\StringTok{ }\NormalTok{theta[}\DecValTok{1}\OperatorTok{:}\NormalTok{p]}
\NormalTok{  tau_param <-}\StringTok{ }\NormalTok{theta[(p}\OperatorTok{+}\DecValTok{1}\NormalTok{)}\OperatorTok{:}\NormalTok{(p}\OperatorTok{+}\NormalTok{n}\OperatorTok{*}\NormalTok{nrandom)]}
\NormalTok{  xi_param <-}\StringTok{ }\NormalTok{theta[(p}\OperatorTok{+}\NormalTok{n}\OperatorTok{*}\NormalTok{nrandom}\OperatorTok{+}\DecValTok{1}\NormalTok{)}\OperatorTok{:}\NormalTok{(p}\OperatorTok{+}\NormalTok{n}\OperatorTok{*}\NormalTok{nrandom}\OperatorTok{+}\NormalTok{nrandom)]}
\NormalTok{  Dhalf <-}\StringTok{ }\KeywordTok{diag}\NormalTok{(}\KeywordTok{exp}\NormalTok{(xi_param), nrandom, nrandom)}

\NormalTok{  L <-}\StringTok{ }\KeywordTok{diag}\NormalTok{(nrandom)}
\NormalTok{  LDhalf <-}\StringTok{ }\NormalTok{L }\OperatorTok{%*%}\StringTok{ }\NormalTok{Dhalf}
\NormalTok{  LDhalf_block <-}\StringTok{ }\KeywordTok{kronecker}\NormalTok{(}\KeywordTok{diag}\NormalTok{(n), LDhalf)}
\NormalTok{  u_param <-}\StringTok{ }\NormalTok{LDhalf_block }\OperatorTok{%*%}\StringTok{ }\NormalTok{tau_param}
\NormalTok{  XZbetau <-}\StringTok{ }\NormalTok{X }\OperatorTok{%*%}\StringTok{ }\NormalTok{beta_param }\OperatorTok{+}\StringTok{ }\NormalTok{Z }\OperatorTok{%*%}\StringTok{ }\NormalTok{u_param}

\NormalTok{  onev <-}\StringTok{ }\KeywordTok{rep}\NormalTok{(}\DecValTok{1}\NormalTok{, }\KeywordTok{length}\NormalTok{(y))}
\NormalTok{  log_likelihood <-}\StringTok{ }\OperatorTok{-}\KeywordTok{t}\NormalTok{(onev) }\OperatorTok{%*%}\StringTok{ }\KeywordTok{exp}\NormalTok{(XZbetau) }\OperatorTok{+}\StringTok{ }\NormalTok{y }\OperatorTok{%*%}\StringTok{ }\NormalTok{XZbetau}
\NormalTok{  log_beta_prior <-}\StringTok{ }\OperatorTok{-}\StringTok{ }\DecValTok{1}\OperatorTok{/}\DecValTok{2}\OperatorTok{*}\KeywordTok{t}\NormalTok{(beta_param)}\OperatorTok{%*%}\StringTok{ }\NormalTok{beta_param}\OperatorTok{/}\NormalTok{sig2beta}
\NormalTok{  log_tau_prior <-}\StringTok{ }\DecValTok{-1}\OperatorTok{/}\DecValTok{2} \OperatorTok{*}\StringTok{ }\KeywordTok{t}\NormalTok{(tau_param) }\OperatorTok{%*%}\StringTok{ }\NormalTok{tau_param}
\NormalTok{  log_xi_prior <-}\StringTok{ }\OperatorTok{-}\NormalTok{(nuxi }\OperatorTok{+}\StringTok{ }\DecValTok{1}\NormalTok{)}\OperatorTok{/}\DecValTok{2} \OperatorTok{*}\StringTok{ }\KeywordTok{log}\NormalTok{(}\DecValTok{1} \OperatorTok{+}\StringTok{ }\DecValTok{1}\OperatorTok{/}\NormalTok{nuxi }\OperatorTok{*}\StringTok{ }
\StringTok{                                        }\KeywordTok{exp}\NormalTok{(}\DecValTok{2}\OperatorTok{*}\NormalTok{xi_param) }\OperatorTok{/}\StringTok{ }\NormalTok{Axi}\OperatorTok{^}\DecValTok{2}\NormalTok{)}

\NormalTok{  result <-}\StringTok{ }\NormalTok{log_likelihood }\OperatorTok{+}\StringTok{ }\NormalTok{log_beta_prior }\OperatorTok{+}\StringTok{ }\NormalTok{log_tau_prior }\OperatorTok{+}\StringTok{ }
\StringTok{    }\KeywordTok{sum}\NormalTok{(log_xi_prior)}
  \KeywordTok{return}\NormalTok{(}\KeywordTok{as.numeric}\NormalTok{(result))}
\NormalTok{\}}
\end{Highlighting}
\end{Shaded}

\begin{Shaded}
\begin{Highlighting}[]
\NormalTok{g_glmm_poisson_posterior <-}\StringTok{ }\ControlFlowTok{function}\NormalTok{(theta, y, X, Z, n, }\DataTypeTok{nrandom=}\DecValTok{1}\NormalTok{,}
                                     \DataTypeTok{nuxi=}\DecValTok{1}\NormalTok{, }\DataTypeTok{Axi=}\DecValTok{25}\NormalTok{, }\DataTypeTok{sig2beta=}\FloatTok{1e3}\NormalTok{) \{}
\NormalTok{  Z <-}\StringTok{ }\KeywordTok{as.matrix}\NormalTok{(Z)}
\NormalTok{  p <-}\StringTok{ }\KeywordTok{ncol}\NormalTok{(X)}

\NormalTok{  beta_param <-}\StringTok{ }\NormalTok{theta[}\DecValTok{1}\OperatorTok{:}\NormalTok{p]}
\NormalTok{  tau_param <-}\StringTok{ }\NormalTok{theta[(p}\OperatorTok{+}\DecValTok{1}\NormalTok{)}\OperatorTok{:}\NormalTok{(p}\OperatorTok{+}\NormalTok{n}\OperatorTok{*}\NormalTok{nrandom)]}

\NormalTok{  xi_param <-}\StringTok{ }\NormalTok{theta[(p}\OperatorTok{+}\NormalTok{n}\OperatorTok{*}\NormalTok{nrandom}\OperatorTok{+}\DecValTok{1}\NormalTok{)}\OperatorTok{:}\NormalTok{(p}\OperatorTok{+}\NormalTok{n}\OperatorTok{*}\NormalTok{nrandom}\OperatorTok{+}\NormalTok{nrandom)]}
\NormalTok{  Dhalf <-}\StringTok{ }\KeywordTok{diag}\NormalTok{(}\KeywordTok{exp}\NormalTok{(xi_param), nrandom, nrandom)}

\NormalTok{  L <-}\StringTok{ }\KeywordTok{diag}\NormalTok{(nrandom)}

\NormalTok{  LDhalf <-}\StringTok{ }\NormalTok{L }\OperatorTok{%*%}\StringTok{ }\NormalTok{Dhalf}
\NormalTok{  LDhalf_block <-}\StringTok{ }\KeywordTok{kronecker}\NormalTok{(}\KeywordTok{diag}\NormalTok{(n), LDhalf)}
\NormalTok{  u_param <-}\StringTok{ }\NormalTok{LDhalf_block }\OperatorTok{%*%}\StringTok{ }\NormalTok{tau_param}

\NormalTok{  XZbetau <-}\StringTok{ }\NormalTok{X }\OperatorTok{%*%}\StringTok{ }\NormalTok{beta_param }\OperatorTok{+}\StringTok{ }\NormalTok{Z }\OperatorTok{%*%}\StringTok{ }\NormalTok{u_param}

\NormalTok{  L_block <-}\StringTok{ }\KeywordTok{kronecker}\NormalTok{(}\KeywordTok{diag}\NormalTok{(n), L)}
\NormalTok{  Dhalf_block <-}\StringTok{ }\KeywordTok{kronecker}\NormalTok{(}\KeywordTok{diag}\NormalTok{(n), Dhalf)}

\NormalTok{  g_beta <-}\StringTok{ }\KeywordTok{t}\NormalTok{(X) }\OperatorTok{%*%}\StringTok{ }\NormalTok{(}\OperatorTok{-}\KeywordTok{exp}\NormalTok{(XZbetau) }\OperatorTok{+}\StringTok{ }\NormalTok{y) }\OperatorTok{-}\StringTok{ }
\StringTok{    }\NormalTok{(beta_param)}\OperatorTok{/}\NormalTok{sig2beta}
\NormalTok{  g_tau <-}\StringTok{ }\KeywordTok{t}\NormalTok{(LDhalf_block) }\OperatorTok{%*%}\StringTok{ }\KeywordTok{t}\NormalTok{(Z) }\OperatorTok{%*%}\StringTok{ }
\StringTok{    }\NormalTok{(}\OperatorTok{-}\KeywordTok{exp}\NormalTok{(XZbetau) }\OperatorTok{+}\StringTok{ }\NormalTok{y) }\OperatorTok{-}\StringTok{ }\NormalTok{tau_param}
\NormalTok{  zero_v <-}\StringTok{ }\KeywordTok{rep}\NormalTok{(}\DecValTok{0}\NormalTok{, nrandom)}
\NormalTok{  g_xi <-}\StringTok{ }\KeywordTok{sapply}\NormalTok{(}\KeywordTok{seq_along}\NormalTok{(}\DecValTok{1}\OperatorTok{:}\NormalTok{nrandom), }\ControlFlowTok{function}\NormalTok{(jj) \{}
\NormalTok{    zv <-}\StringTok{ }\NormalTok{zero_v}
\NormalTok{    zv[jj] <-}\StringTok{ }\DecValTok{1}
\NormalTok{    bd <-}\StringTok{ }\KeywordTok{kronecker}\NormalTok{(}\KeywordTok{diag}\NormalTok{(n), }\KeywordTok{diag}\NormalTok{(zv, nrandom, nrandom))}
      \KeywordTok{t}\NormalTok{(L_block }\OperatorTok{%*%}\StringTok{ }\NormalTok{bd }\OperatorTok{%*%}\StringTok{ }\NormalTok{Dhalf_block }\OperatorTok{%*%}\StringTok{ }\NormalTok{tau_param) }\OperatorTok{%*%}\StringTok{ }
\StringTok{        }\KeywordTok{t}\NormalTok{(Z) }\OperatorTok{%*%}\StringTok{ }\NormalTok{(}\OperatorTok{-}\KeywordTok{exp}\NormalTok{(XZbetau) }\OperatorTok{+}\StringTok{ }\NormalTok{y)}
\NormalTok{  \})}
\NormalTok{  g_xi <-}\StringTok{ }\NormalTok{g_xi }\OperatorTok{-}\StringTok{ }\NormalTok{(nuxi }\OperatorTok{+}\StringTok{ }\DecValTok{1}\NormalTok{) }\OperatorTok{/}\StringTok{ }\NormalTok{(}\DecValTok{1} \OperatorTok{+}\StringTok{ }\NormalTok{nuxi}\OperatorTok{*}\NormalTok{Axi}\OperatorTok{^}\DecValTok{2} \OperatorTok{*}\StringTok{ }
\StringTok{                                 }\KeywordTok{exp}\NormalTok{(}\OperatorTok{-}\DecValTok{2}\OperatorTok{*}\NormalTok{xi_param)) }\OperatorTok{+}\StringTok{ }\DecValTok{1}

\NormalTok{  g_all <-}\StringTok{ }\KeywordTok{c}\NormalTok{(}\KeywordTok{as.numeric}\NormalTok{(g_beta),}
             \KeywordTok{as.numeric}\NormalTok{(g_tau),}
\NormalTok{             g_xi)}

  \KeywordTok{return}\NormalTok{(g_all)}
\NormalTok{\}}
\end{Highlighting}
\end{Shaded}

With the dependent variable and design matrices defined, we run HMC for the Poisson mixed effects model. Initial values are set to zero and default hyperparameters are selected.

\begin{CodeChunk}
\begin{CodeInput}
R> N <- 2e3
R> 
R> set.seed(412)
R> initvals <- c(rep(0, 4), rep(0, 10), 0)
R> eps_vals <- c(3e-2, 3e-2, 3e-2, 1e-3, rep(1e-1, 10), 3e-2)
R> 
R> fm3_hmc <- hmc(N = N, theta.init = initvals, epsilon = eps_vals, L = 10, 
		logPOSTERIOR = glmm_poisson_posterior, 
		glogPOSTERIOR = g_glmm_poisson_posterior, 
		varnames = c(colnames(X), paste0("tau", 1:ncol(Z)), "xi"), 
		param = list(y = y, X = X, Z = Z, n = 10), 
		chains = 2, parallel = FALSE)
R> 
R> fm3_hmc$accept / N
\end{CodeInput}

\begin{CodeOutput}
[1] 0.8005 0.8045
\end{CodeOutput}
\end{CodeChunk}

The acceptance rate for this model is 0.80 for both chains, which is within the range of efficiently tuned HMC applications. From the \textit{summary} function, we note that the $\hat{R}$ statistics are all close to one, indicating that both chains converged to the same distribution. 

\begin{CodeChunk}
\begin{CodeInput}
R> summary(fm3_hmc, burnin=200)
\end{CodeInput}
\end{CodeChunk}

\resizebox{1 \textwidth}{!}{
\begin{tabular}{lrrrrrrrr}
\toprule
  & 2.5\% & 5\% & 25\% & 50\% & 75\% & 95\% & 97.5\% & rhat\\
\midrule
intercept & -1.130 & -0.955 & -0.390 & -0.088 & 0.178 & 0.555 & 0.693 & 1.000\\
factor.year.2005 & -1.372 & -1.260 & -0.906 & -0.667 & -0.432 & -0.100 & -0.004 & 1.000\\
factor.year.2006 & -1.013 & -0.924 & -0.591 & -0.383 & -0.164 & 0.120 & 0.215 & 1.001\\
prev & 0.006 & 0.010 & 0.018 & 0.023 & 0.028 & 0.037 & 0.040 & 1.000\\
tau1 & -2.411 & -2.135 & -1.319 & -0.780 & -0.246 & 0.581 & 0.817 & 1.000\\
tau2 & -1.730 & -1.421 & -0.687 & -0.184 & 0.296 & 1.019 & 1.221 & 1.000\\
tau3 & -1.983 & -1.755 & -1.011 & -0.569 & -0.081 & 0.584 & 0.807 & 1.000\\
tau4 & -0.568 & -0.383 & 0.308 & 0.731 & 1.220 & 1.915 & 2.176 & 1.002\\
tau5 & -1.610 & -1.342 & -0.560 & -0.107 & 0.307 & 0.974 & 1.196 & 1.000\\
tau6 & -0.560 & -0.183 & 0.577 & 1.112 & 1.550 & 2.268 & 2.487 & 1.000\\
tau7 & -1.047 & -0.835 & -0.175 & 0.236 & 0.642 & 1.227 & 1.464 & 1.000\\
tau8 & -1.660 & -1.389 & -0.654 & -0.190 & 0.232 & 0.874 & 1.090 & 1.001\\
tau9 & -0.752 & -0.391 & 0.427 & 0.923 & 1.353 & 2.053 & 2.313 & 1.001\\
tau10 & -2.504 & -2.258 & -1.500 & -0.996 & -0.485 & 0.222 & 0.489 & 1.000\\
xi & -2.588 & -2.101 & -0.781 & -0.363 & -0.071 & 0.402 & 0.554 & 1.003\\
\bottomrule
\end{tabular}
}
\\
\\
In this example, the posterior estimates are comparable to frequentist estimates. We use the \texttt{lme4} package \citep{bates2007lme4} to provide frequentist parameter estimates as a comparison to HMC.

\begin{CodeChunk}
\begin{CodeInput}
fm3 <- glmer(shells ~ prev + factor(year) + (1 | Site), 
	family = poisson, data = Gdat, 
	control = glmerControl(optimizer = "bobyqa", 
			check.conv.grad = .makeCC("warning", 0.05)))
\end{CodeInput}
\end{CodeChunk}

\begin{CodeChunk}
\begin{CodeInput}
R> coef(summary(fm3))
\end{CodeInput}
\end{CodeChunk}

\begin{tabular}{lrrrr}
\toprule
  & Estimate & Std. Error & z value & Pr($>\lvert z \rvert$)\\
\midrule
(Intercept) & -0.058 & 0.397 & -0.145 & 0.884\\
prev & 0.022 & 0.008 & 2.891 & 0.004\\
factor(year)2005 & -0.654 & 0.357 & -1.830 & 0.067\\
factor(year)2006 & -0.374 & 0.323 & -1.157 & 0.247\\
\bottomrule
\end{tabular}
\\
\\
Next, we store the frequentist fixed effects estimates in R variables.

\begin{CodeChunk}
\begin{CodeInput}
R> freqvals_fixed <- c(fixef(fm3))
R> freqvals_fixed <- freqvals_fixed[c(1, 3, 4, 2)]
\end{CodeInput}
\end{CodeChunk}

We also compare the random effects parameter estimates with \textbf{lme4}. We apply the linear transformation back to \(\mathbf{u}\) for comparison.

\begin{CodeChunk}
\begin{CodeInput}
R> u.freq <- ranef(fm3)$Site[, 1]
R> lambda.freq <- sqrt(VarCorr(fm3)$Site[1])
R> 
R> fm3_hmc$thetaCombined <- lapply(fm3_hmc$thetaCombined, function(xx) {
		tau_mx <- as.matrix(xx[, grepl("tau", colnames(xx))])
		u_mx <- tau_mx * exp(xx[, "xi"])
		u_df <- as.data.frame(u_mx)
		colnames(u_df) <- paste0("u", 1:ncol(u_df))
		xx <- cbind(xx, u_df, exp(xx[, "xi"))
		colnames(xx)[ncol(xx)] <- "lambda"
		xx
	})
\end{CodeInput}
\end{CodeChunk}

\begin{figure}
\begin{center}
\includegraphics[width=0.7\textwidth]{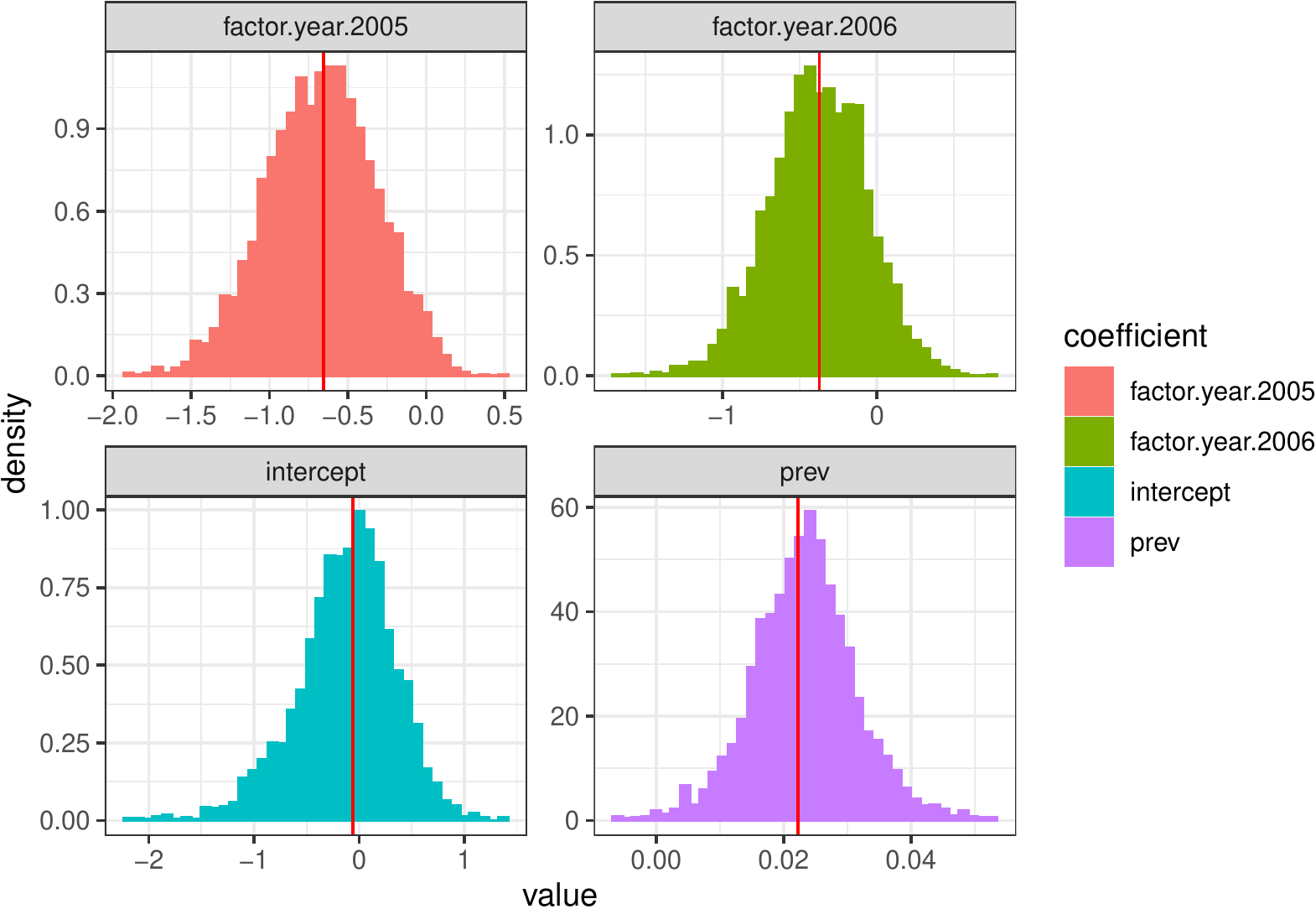}
\end{center}
\caption{Diagnostic plots comparing HMC (histograms) to frequentist (vertical lines) estimates of fixed effects parameters for example 3. The HMC estimates are in line with those from \textit{glmer}. }
\label{fig:diag3fixed}
\end{figure}

\begin{figure}
\begin{center}
\includegraphics[width=0.7\textwidth]{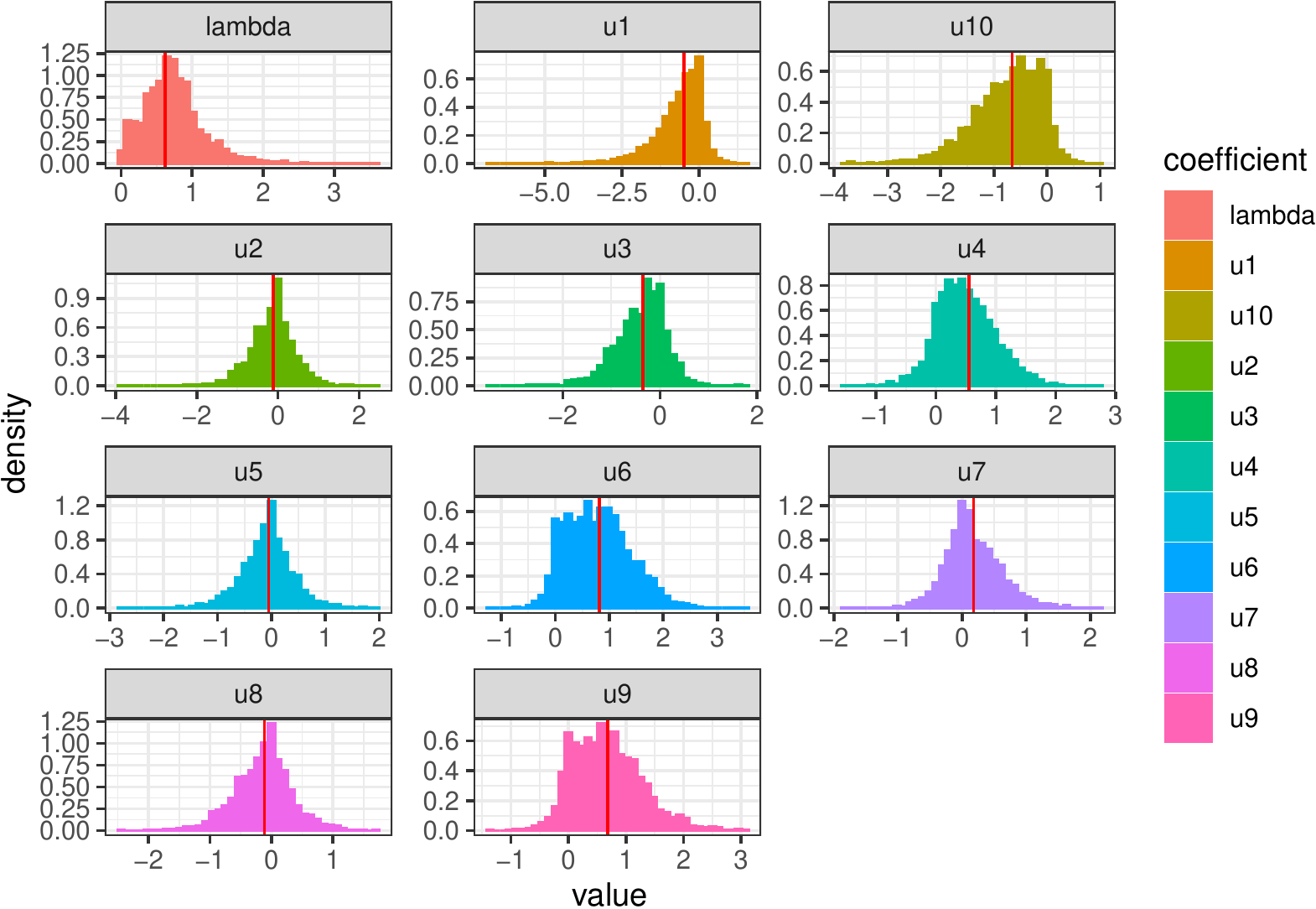}
\end{center}
\caption{Diagnostic plots comparing HMC (histograms) to frequentist (vertical lines) estimates of random effects parameters for example 3. The HMC estimates are in line with those from \textit{glmer}. }
\label{fig:diag3random}
\end{figure}

The frequentist estimates for fixed effects and random effects are close to HMC estimates in this example. See Fig \ref{fig:diag3fixed} and \ref{fig:diag3random}.

\begin{CodeChunk}
\begin{CodeInput}
R> diagplots(fm3_hmc, burnin = 200, compaison.theta = freqvals_fixed, cols = 1:4)
\end{CodeInput}
\end{CodeChunk}

\begin{CodeChunk}
\begin{CodeInput}
R> diagplots(fm3_hmc, burnin = 200, 
	comparison.theta = c(u.freq, lambda.freq), cols = 16:26)
\end{CodeInput}
\end{CodeChunk}

\newpage
\bibliographystyle{tfcad}
\bibliography{hmc.bib}